\documentclass[11pt]{iopart2}

\usepackage{graphicx,bm,float,amsmath,amssymb,tikz}

\usepackage[normalem]{ulem}

\makeatletter

\usepackage[english]{babel}
\usepackage[T1]{fontenc}
\usepackage[utf8]{inputenc}

\begin{document}

\title{Multicomponent gauge-Higgs models with discrete Abelian gauge groups}

\author{Giacomo Bracci Testasecca$^1$ and Andrea Pelissetto$^{1,2}$} 
\address{$^1$Dipartimento di Fisica, Universit\`a di Roma ``La Sapienza",
        P.le Aldo Moro 2, I-00185 Roma, Italy}
\address{$^2$ INFN, Sezione di Roma,
        P.le Aldo Moro 2, I-00185 Roma, Italy}

\date{\today}

\begin{abstract}
We consider a variant of the charge-$Q$ compact Abelian-Higgs model, in which 
an $N_f$-dimensional complex vector is coupled with an Abelian ${\mathbb Z}_q$
gauge field. For $N_f=2$ and $Q=1$ we observe several transition lines 
that belong to the O(4), O(3), and O(2) vector universality classes, 
depending on the symmetry breaking pattern
at the transition. The universality class is independent of $q$ as long 
as $q\ge 3$. The universality class of the transition is uniquely determined 
by the behavior of the scalar fields; gauge fields do not play any role. 
We also investigate the system for $N_f=15$ and $Q=2$. In the presence 
of U(1) gauge fields, the system undergoes transitions associated with 
charged fixed points of the Abelian-Higgs field theory. These continuous
transitions turn into first-order ones when the U(1) gauge fields are replaced 
by the discrete ${\mathbb Z}_q$ fields: in the present 
compact model charged transitions appear to be very
sensitive to the nature of the gauge fields.
\end{abstract}



\section{Introduction}
\label{intro}

Symmetries have always played a central role in the development of physical
models. Global symmetries are commonly used to characterize the different
phases of matter \cite{Anderson-book} and to understand the nature and 
properties of the transition lines between them. In the absence of 
local symmetries, the universal features of critical transitions can 
be understood using the traditional 
Landau-Ginzburg-Wilson (LGW) approach to critical phenomena
\cite{Landau-book, WK-74, Fisher-75, PV-02,ZJ-book}. It assumes that 
universal properties can be explained by a 
statistical field theory, in which a local order parameter plays the 
role of fundamental field. The universality classes are uniquely
determined by the breaking pattern of the global symmetry, i.e., by the 
symmetry of the two phases coexisting at the transition, and by the
transformation properties of the order parameter under the symmetry group.

Local gauge symmetries also play a
fundamental role, both in high-energy physics \cite{Weinberg-book}, 
and in condensed-matter
physics, where their applications span from superconductivity
\cite{Anderson-63} to topological order and quantum phase transitions,
see, e.g., Refs.~\cite{Fradkin_book,MM_book,Sachdev-19}.
The critical behavior of three-dimensional 
systems with global and local symmetries is, at 
present, not fully understood. First of all, in the presence of gauge degrees
of freedom, one may have topologically ordered phases, that cannot be
characterized in terms of local order parameters, whose existence is 
the basic tenet of the LGW approach
\cite{Wegner-71,Kogut-79,SBSVF-04, WNMXS-17}. Second, it is unclear under which
conditions the gauge degrees of freedom play a role at transitions where 
global symmetries are broken and matter fields show long-range correlations.

Critical transitions in three-dimensional 
lattice gauge theories can be
classified in three different groups. First, there are topological transitions,
where only the gauge degrees of freedom play a role and 
the global symmetry of the
coexisting phases is the same. Matter fields are insensitive to the behavior 
of the gauge degrees of freedom and the critical behavior of the gauge modes is 
the same as in pure gauge models. Transitions of this type have been observed,
e.g., in the three-dimensional Abelian Higgs (AH) model, both with 
noncompact fields \cite{BPV-21-NCQED} and with charge-$Q$ scalar fields
\cite{BPV-20-AHq2,BPV-22-AHq}.

Second, one should consider transitions in which only matter correlations are
critical. At the transition gauge modes do not display long-range correlations,
but are still crucial in determining the universality class of the critical 
behavior. 
Indeed, the gauge symmetry prevents non-gauge-invariant matter observables
from developing long-range order: the gauge symmetry hinders
some matter degrees of freedom---those that are not gauge
invariant---from becoming critical. In this case, the gauge symmetry 
plays a crucial role in determining the order parameter of the transition
and its transformation properties under the global symmetry group. 
However, in the absence of long-range gauge modes, one can still use
the LGW approach, without including the
gauge fields,  to predict the critical behavior. 
The lattice CP$^{N-1}$ model is an example of a U(1)
gauge model that shows this type of behavior
~\cite{PV-19-CP,PV-20-largeNCP}. Transitions of this type also occur in several 
non-Abelian gauge models 
~\cite{BPV-19-sqcd,BFPV-21-sunfu,BFPV-21-sunadj,BPV-20-son}. 
The finite-temperature transition of massless quantum chromodynamics has also 
been assumed to have these features \cite{PW-84}.

Finally, there are transitions at which both matter 
and gauge correlations are critical. In this case
an appropriate effective field-theory description of the critical
behavior requires the inclusion of explicit gauge fields.
Transitions of this type have been observed in very few cases; 
for instance, in
the lattice AH model with noncompact gauge
fields~\cite{BPV-21-NCQED} (along the transition line that separates the
Coulomb and the Higgs phase), and
in the lattice AH model with compact gauge fields and 
scalar fields with charge $Q\ge 2$~\cite{BPV-22-AHq} 
(along the  transition line between
the confined and the deconfined phase), when the number of components 
is sufficiently large (numerical simulations find the charged transition
for $N_f\ge 10$). In the non-Abelian case, at present,
there is some evidence of this type of behavior only 
in SU(2) gauge theories coupled with 
scalar matter in the fundamental representation \cite{BFPV-21-sunfu}.

In this work we consider charge-$Q$ $N_f$-component scalar fields interacting 
by means of a U($N_f$)-invariant Hamiltonian 
and couple them with ${\mathbb Z}_q$
compact gauge fields.  The model represents a 
generalization of the compact charge-$Q$  AH 
model \cite{FS-79,BPV-22-AHq}, in which
the gauge group U(1) is replaced by its subgroup ${\mathbb Z}_q$. We 
determine the phase diagram of the model and the nature of the critical lines, 
with the purpose of verifying whether 
the previous considerations also apply to systems with discrete gauge groups.

We will study the model for two different values of the charge and number of
components, which have been chosen on the basis of the 
results for U(1) gauge models. 
First, we perform a numerical study of the model for $N_f=2$ and 
$Q=1$. In the presence of compact U(1)
gauge fields (it represents the standard compact lattice AH model), 
it shows a single critical line associated with the
breaking of the global SU(2) symmetry \cite{PV-19-AH3d}. The transitions 
have an effective LGW description that predicts O(3) vector behavior. In the 
LGW approach to gauge systems, the nature of the gauge group is not relevant,
and thus we expect the same critical behavior in ${\mathbb Z}_q$ gauge models, 
provided the breaking pattern of the global symmetry is the same. Our numerical
results for $q=3$ and $q=4$ are in full agreement with the LGW predictions: 
also for these values of $q$, transitions associated with the breaking of the
SU(2) global symmetry belong to the O(3) vector universality class. 
In the discrete gauge model, the global symmetry group is larger. The model 
is also invariant under U(1)$/{\mathbb Z}_q$ global transformations. 
For $q=3$ and 4, 
we find a second transition line where this symmetry breaks. 
Also along this transition line the gauge 
group does not play any role: the transitions belong to the O(2)/XY 
vector universality class, as expected on the basis of the LGW analysis.

The second case of interest corresponds to systems with $N_f=15$ and $Q=2$. 
The corresponding U(1) gauge-invariant model was studied in 
Ref.~\cite{BPV-22-AHq}, finding a {\em charged} transition line, where both 
scalar and gauge degrees of freedom play a role. Gauge fields must be included 
in the effective description of the critical behavior and indeed the observed 
behavior is consisted with that predicted by the AH field
theory \cite{HLM-74,FH-96,IZMHS-19,YKK-96,MZ-03}. A natural question is whether
the {\em charged} transition survives if one replaces the continuous 
gauge group U(1) with the discrete ${\mathbb Z}_q$ group.
In the case of a global 
U(1) symmetry, it is well known that systems with microscopic ${\mathbb Z}_q$
symmetry may have transitions in the U(1)/O(2) universality class if $q\ge 4$
(see Refs.~\cite{HS-03,SGS-20} and references therein). 
At the transition one observes an enlargement of the global 
symmetry: The large-distance
behavior is invariant under transformations that are not symmetries of the 
microscopic theory. Here, we investigate whether a similar 
phenomenon occurs for gauge symmetries, i.e., whether it is possible 
to have a {\em gauge symmetry enlargement}. Simulations with $q=6$ and 10
indicate that no such symmetry enlargement occurs. We identify a transition
line that separates two phases that have the same features as those 
coexisting along the charged transition line in the U(1) gauge model, but 
in ${\mathbb Z}_q$ gauge compact models these transitions 
turn out to be of first order for both values of $q$. Apparently, the
microscopic model should be exactly U(1) gauge invariant to allow the system to
develop critical charged transitions.

The paper is organized as follows. 
In Sec.~\ref{sec2} we define the model, while in Sec.~\ref{sec3} 
we specify the quantities that are determined in the Monte Carlo simulations.
Some limiting cases, useful to understand the general featurs of the 
phase diagram,  are discussed in Sec.~\ref{sec4}, while 
the numerical results are presented in Sec.~\ref{sec5} (results for $N_f=2$)
and in Sec.~\ref{sec6} (results for $N_f=15$).
Finally, in Sec.~\ref{sec7} we summarize the results and draw our conclusions. 

\section{The model} \label{sec2}

We consider a ${\mathbb Z}_q$ gauge model coupled with an 
$N_f$-component charge-$Q$ complex scalar field 
defined on a cubic lattice. 
The fundamental
fields are complex $N_f$-dimensional vectors 
${\bm w}_{\bm x}$, satisfying $|w_{\bm x}| = 1$,
associated with the sites of the lattice,
and phases $\sigma_{{\bm x},\mu}$, 
$|\sigma_{{\bm x},\mu}|=1$, associated with the lattice
links. The gauge variables can only take $q$ values.
More precisely, we set
\begin{equation}
\sigma = \exp (2 \pi i n /q),
\label{defwsigma}
\end{equation}
where $n = 0,\ldots, q-1$.

The Hamiltonian is 
\begin{equation}
    H = H_{\rm kin} + H_{\rm g}.
\label{Hamiltonian}
\end{equation}
The first term is 
\begin{equation}
H_{\rm kin} = - J\ \hbox{Re} \sum_{{\bm x}, \mu}
\left( \overline{\bm w}_{\bm x} \cdot {\bm w}_{{\bm x}+\hat\mu}
\sigma_{{\bm x},\mu}^Q \right) ,
\label{hcpnla}
\end{equation}
where the sum is over all lattice sites ${\bm x}$ and directions $\mu$
($\hat{\mu}$ are the corresponding unit vectors), and $Q$ is the 
integer charge of the scalar field. Since $\sigma_{{\bm x},\mu}^q = 1$,
we can limit ourselves to charges $Q$ satisfying $1 \le Q \le q/2$. 

The second term is 
\begin{equation}
H_{\rm g} = - g \sum_{{\bm x},\mu>\nu} \hbox{Re }\Pi_{{\bm x},\mu\nu},
\end{equation}
where the sum is over all lattice plaquettes, and the plaquette 
contribution is given by
\begin{equation}
\Pi_{{\bm x},\mu\nu} = 
   \sigma_{{\bm x},\mu} \sigma_{{\bm x} + \hat{\mu},\nu}
   \overline{\sigma}_{{\bm x} + \hat{\nu},\mu} \overline{\sigma}_{{\bm x},\nu}, 
\end{equation}
The partition function is 
\begin{equation}
Z = \sum_{\{w,\sigma\}} e^{-H/T}.
\label{Z-model}
\end{equation}
In the following we will use $\beta = J/T$ and $\kappa = g/T$ as 
independent variables.

The model is invariant under  
the local ${\mathbb Z}_q$ transformations 
\begin{equation}
{\bm w}_{\bm x} \to \lambda_{\bm x}^Q {\bm w}_{\bm x} \qquad
\sigma_{{\bm x},\mu} \to \lambda_{\bm x} \sigma_{{\bm x},\mu}
\overline{\lambda}_{{\bm x} +\hat{\mu}},
\end{equation}
where $\lambda_{\bm x}$ are phases satisfying $\lambda_{\bm x}^q = 1$. 
It is also invariant under the global U($N_f$) transformations
\begin{equation}
   {\bm w}_{\bm x}  \to V {\bm w}_{\bm x},
\end{equation}
where $V\in U(N_f)$. 

For $q/Q = 2$,
the global symmetry is larger. Indeed, in this case 
$\sigma_{{\bm x},\mu}^Q$ is real (it takes the values $\pm 1$) and this allows
us to rewrite the Hamiltonian as a vector Hamiltonian. We define 
a real field $\Phi^A$ with $2 N_f$ components
\begin{equation}
\Phi^a_{\bm x} = \hbox{Re } w^a_{\bm x} \qquad
\Phi^{a+N_f}_{\bm x} = \hbox{Im } w^a_{\bm x},
\label{def-Phi}
\end{equation}
where $1\le a \le N_f$. Since 
\begin{equation}
 \hbox{Re } \overline{\bm w}_{\bm x} \cdot {\bm w}_{\bm y} = 
   \Phi_{\bm x} \cdot \Phi_{\bm y},
\label{mappingO2N}
\end{equation}
the scalar Hamiltonian can be rewritten as 
\begin{equation}
     H_{\rm kin} = - J \sum_{x\mu} \sigma_{x,\mu}^Q 
    \Phi_{\bm x} \cdot \Phi_{{\bm x} + \hat{\mu}}.
\end{equation}
This Hamiltonian is invariant under the transformations 
$\Phi_{\bm x} \to V \Phi_{\bm x}$, where $V$ is an $O(2N_f)$ matrix. 
For $Q=1$ and $q=2$, the scalar Hamiltonian corresponds to a particular
RP$^{N-1}$ vector model ($N=2N_f$), in which the vector fields take values in
the RP$^{N-1}$ manifold. For $N=3$ this model is relevant for liquid crystals
\cite{dGP-93,LL-73}
and it has been studied, also in the presence of gauge interactions, 
in Refs.~\cite{LRT-93,LNNSWZ-15}.

\section{The observables} \label{sec3}

We simulate the system using a combination of standard Metropolis updates
of the scalar and gauge fields and of 
microcanonical updates of the scalar field.
We compute the energy densities and the specific heats 
\begin{eqnarray}
&& E_k = - {1\over VJ} \langle H_{\rm kin} \rangle,\qquad
C_k ={1\over V J^2}
\left( \langle H^2_{\rm kin} \rangle 
- \langle H_{\rm kin} \rangle^2\right), 
\nonumber  \\
&& E_g = - {1\over Vg} \langle H_{\rm g} \rangle,\qquad
C_g ={1\over V g^2}
\left( \langle H^2_{\rm g} \rangle 
- \langle H_{\rm g} \rangle^2\right), 
\label{ecvdef}
\end{eqnarray}
where $V=L^3$.

We consider the two-point correlation function of the gauge-invariant 
tensor combination
\begin{equation}
    T^{ab} = \overline{w}^a w^b - {1\over N_f} \delta^{ab},
\end{equation}
defined as 
\begin{equation}
G_T({\bm x},{\bm y}) = 
   \sum_{ab} 
   \langle T^{ab}_{\bm x} T^{ba}_{\bm y} \rangle.
\label{gT} 
\end{equation}
The correlation function is invariant under local ${\mathbb Z}_q$ 
transformations and global U($N_f$) transformations.
We also consider the scalar-field combination
\begin{equation}
\Sigma_k^{a_1,\ldots,a_k} = w^{a_1}\ldots w^{a_k}.
\end{equation}
If $k Q$ is a multiple of $q$, this quantity is gauge invariant.
Correspondingly, we consider the charge-$k$ correlation function
\begin{eqnarray}
G_k({\bm x},{\bm y}) &= &
   \hbox{Re} \, \langle \overline{\Sigma}_{k,\bm x} \cdot {\Sigma}_{k,\bm y} 
   \rangle = 
   \hbox{Re} \, 
   \langle (\overline{\bm w}_{\bm x}\cdot {\bm w}_{\bm y})^k \rangle.
\label{Gq} 
\nonumber 
\end{eqnarray}
The two local order parameters $T^{ab}_{\bm x}$ and 
$\Sigma_{{\bm x},k}^{a_1,\ldots,a_k}$ belong to two different
representations of the global symmetry group U($N_f$). 
In particular, $T^{ab}_{\bm x}$ is invariant under transformations 
belonging to the Abelian subgroup
U(1), while $\Sigma_{{\bm x}k}^{a_1,\ldots,a_k}$
transforms nontrivially. For $q=2Q$, the symmetry group is O($2N_f$). 
In this case,
the gauge-invariant order parameters $T^{ab}$ and $\Sigma_{2}^{ab}$ are 
equivalent, being both related to 
\begin{equation}
Q^{AB} = \Phi^A \Phi^B - {1\over 2 N_f} \delta^{AB},
\end{equation}
where $\Phi$ is defined in Eq.~(\ref{def-Phi}). 
If 
\begin{equation} 
G_\Phi({\bm x,\bm y}) = 
   \sum_{AB} \langle Q_{\bm x}^{AB} Q_{\bm y}^{BA} \rangle,
\end{equation}
one easily derives 
\begin{eqnarray} 
    G_T({\bm x,\bm y}) &=& {2 (N_f-1)\over 2 N_f - 1} G_\Phi({\bm x,\bm y}),
\nonumber \\
    G_2({\bm x,\bm y}) &=& {2 N_f \over 2 N_f - 1} G_\Phi({\bm x,\bm y}) .
\end{eqnarray}
The calculation of $G_k({\bm x,\bm y})$ is particularly time-consuming 
when $k$ and $N_f$ are large. In some cases, we have used the following 
identity that relies on the $U(N_f)$ invariance of the theory: 
\begin{equation}
G_k({\bm x,\bm y}) = {(k + N_f - 1)!\over k! (N_f-1)!} 
   {\langle ({\bm a}\cdot \overline{\bm w})^k ({\bm b}\cdot {\bm w})^k\rangle
   \over ({\bm a}\cdot {\bm b})^k},
\end{equation}
where ${\bm a}$ and ${\bm b}$ are arbitrary $N_f$-dimensional vectors. 
Choosing vectors ${\bm a} = {\bm b} = (0,\ldots,0,1,0,\ldots)$, we obtain
\begin{equation}
G_k({\bm x,\bm y}) = {(k + N_f - 1)!\over k! N_f!}  \sum_\alpha
   \langle \overline{\bm w}^k_\alpha {\bm w}^k_\alpha\rangle.
\end{equation}
To define the corresponding correlation lengths, we define the 
Fourier transform 
\begin{equation}
\widetilde{G}_\#({\bm p}) = 
    {1\over V} \sum_{{\bm x},{\bm y}} 
      e^{i{\bm p}\cdot ({\bm x} - {\bm y})} G_\#({\bm x},{\bm y})
\end{equation}
($V$ is the volume) of the two correlation functions.
The corresponding susceptibilities and 
correlation lengths are defined as 
\begin{eqnarray}
&&\chi_\# =  
\widetilde{G}_\#({\bm 0}), 
\label{chisusc}\\
&&\xi^2_\# \equiv  {1\over 4 \sin^2 (\pi/L)}
{\widetilde{G}_\#({\bm 0}) - \widetilde{G}_\#({\bm p}_m)\over 
\widetilde{G}_\#({\bm p}_m)},
\label{xidefpb}
\end{eqnarray}
where ${\bm p}_m =
(2\pi/L,0,0)$. Note that, if one uses open boundary conditions, the choice 
of ${\bm p}_m$ is somewhat arbitrary. Other choices, as long as they satisfy 
$|p_m| \sim 1/L$, would be equally valid.

In our finite-size scaling (FSS) analysis 
we use renormalization-group invariant 
quantities. We consider
\begin{equation}
R_{\xi,\#} = \xi_\#/L
\end{equation}
and the Binder parameters
\begin{equation}
U_\# = {\langle \mu_{2,\#}^2\rangle \over \langle \mu_{2,\#}\rangle^2} , \qquad
\mu_{2,\#} = \sum_{{\bm x}{\bm y}}
    G_\#({\bm x},{\bm y}).
\label{binderdef}
\end{equation}
In the disordered phase, we have 
\begin{equation}
U_T = {N^2_f + 1\over N^2_f - 1}, \qquad
U_k = 1 + {k! (N_f-1)!\over (N_f + k - 1)!},
\end{equation}
while in the ordered phase, all these quantities converge to 1. 

To determine the nature of the transition, one can consider the 
$L$ dependence of the 
maximum $C_{\rm max}(L)$ of one of the specific heats.  At a first-order
transition, $C_{\rm max}(L)$ is proportional to the volume $L^3$, while 
at a continuous transition it behaves as 
\begin{equation}
C_{\rm max}(L) = a L^{\alpha/\nu} + C_{\rm reg} .
\end{equation}
The constant term $C_{\rm reg}$, due to the analytic
background, is the dominant contribution if $\alpha < 0$.  The
analysis of the $L$-dependence of $C_{\rm max}(L)$ may allow one 
to distinguish first-order and continuous transitions.
However, experience with
models that undergo weak first-order transitions indicates
that in many cases the analysis of the specific heat is not
conclusive~\cite{CLB-86,VRSB-93}. 
The maximum $C_{\rm max}(L)$  may start scaling as $L^3$ at values of
$L$ that are much larger than those at which simulations can be actually
performed.  A more useful quantity is a Binder parameter $U$.
At first-order transitions, the
maximum $U_{\rm max}(L)$ of $U$ at fixed size $L$ increases with the 
volume~\cite{CLB-86,VRSB-93}. On the other hand, 
$U$ is bounded as $L\to \infty$ at a continuous phase transition.
In this case, in the FSS limit, the Binder parameter as well as 
any renormalization-group invariant quantity $R$ scales as
\begin{equation}
R(\beta,L) \approx  f_R(X) + L^{-\omega} f_{c,R}(X) 
\quad X = (\beta-\beta_c) L^{1/\nu} ,
\label{rsca}
\end{equation}
where $\omega$ is the leading correction-to-scaling exponent. 
Thus, a first-order transition can be identified by verifying that
$U_{\rm max}(L)$ increases with $L$, without the need of explicitly
observing the linear behavior in the volume.

In the case of weak first-order transitions, the nature of the
transition can also be understood from the combined analysis of $U$
and $R_{\xi}$. At a continuous transition, in the FSS
limit any renormalization-group invariant quantity $R$ scales as 
\begin{equation}
R(\beta,L) = F_R(R_\xi) + L^{-\omega} F_{c,R}(R_\xi) + \ldots
\label{r12sca}
\end{equation}
where $F_R(x)$ is universal and $F_{c,R}(x)$ is universal apart from a
multiplicative constant. The Binder parameter $U$ does not obey this scaling
relation at 
first-order transitions, because of the divergence of $U$ for
$L\to\infty$.  Therefore, the absence of data
collapse in plots of $U$ versus $R_\xi$ 
is an early indication of the first-order nature of the
transition.

\section{Some limiting cases} \label{sec4}

The phase diagram of the compact model with
U(1) gauge fields, i.e., in the limit $q\to\infty$,
has been discussed at length in 
Refs.~\cite{FS-79,PV-19-AH3d,BPV-20-AHq2,BPV-22-AHq}, 
while systems with discrete gauge groups are discussed in 
Refs.~\cite{FS-79,BPPW-81,LRT-93,LNNSWZ-15,BPV-22-discrete}.
To derive the phase diagram of the present model, it is useful to 
discuss some limiting cases.

In the limit $\kappa\to\infty$,
the gauge degrees of freedom freeze and one can set 
$\sigma_{{\bm x},\mu} = 1$ on all links (when open boudary conditions
are used, this is also true in a finite volume). Using the 
mapping (\ref{mappingO2N}),
the scalar Hamiltonian becomes  equivalent
to that of the O($2N_f$) vector model,
which undergoes a standard
finite-$\beta$ continuous transition for any $N_f \ge 1$.

For $\beta = 0$, 
there are no scalar fields and one obtains a pure gauge ${\mathbb Z}_q$
model, that can be related by duality~\cite{SSNHS-03} 
to a ${\mathbb Z}_q$ spin model, with 
a global ${\mathbb Z}_q$ symmetry. The ${\mathbb Z}_q$ gauge theory
undergoes a topological transition at $\kappa_c$, which 
belongs to the same universality class as the corresponding transition
in the ${\mathbb Z}_q$ spin clock model. 
For $q=2$ and 4 it belongs to the
Ising universality class, for $q=3$ it is of 
first order, while for $q\ge 5$ the critical behavior is the same as 
in the XY model, see
Refs.~\cite{HS-03,Hasenbusch-19,PSS-20}. 
Estimates of $\kappa_c$ can be found in Ref.~\cite{BCCGPS-14}. 
For $q=2$, we can use duality and 
the results Ref.~\cite{FXL-18} for the standard Ising
model to estimate $\kappa_c = 0.761413292(12)$. For $q\to \infty$, one has
\cite{BCCGPS-14,BPV-22-AHq}
\begin{equation}
\kappa_c\simeq \kappa_{gc} \, q^2,
\end{equation}
where $\kappa_{gc}=0.076051(2)$ is the critical
coupling of the inverted XY model \cite{NRR-03}. 

For $\kappa = 0$, there is no plaquette contribution. We can, first of all,
simplify the scalar-field interaction term, defining
\begin{equation}
    \tau_{{\bm x},\mu} = \sigma_{{\bm x},\mu}^M,
\end{equation}
where $M$ is the greatest common divisor of $Q$ and $q$.
The new field takes $p=q/M$ different values, i.e., it satisfies 
$\tau_{{\bm x},\mu}^p = 1$. If $r = Q/M$, the Hamiltonian becomes 
\begin{equation}
H_{\rm kin} = - J\ \hbox{Re} \sum_{{\bm x}\mu}
   \overline{\bm w}_{\bm x} \cdot {\bm w}_{x+\hat{\mu}} \tau_{{\bm x},\mu}^r.
\label{Hkin_kappa0}
\end{equation} 
Note that $p$ is always larger than 1, since $M\le Q < q$.
Thus, we obtain an effective model for charge-$r$ scalars 
with ${\mathbb Z}_p$ local gauge invariance. 

For $p/r = q/Q = 2$ (in this case we have necessarily $p=2$ and $r=1$),  
the global invariance 
group of the model is the O($2N_f$) group and  one obtains an effective 
RP$^{2N_f-1}$ model.
The critical behavior of RP$^{N-1}$ models is well known: for any $N\ge 3$
they are expected to undergo a first-order transition
\cite{dGP-93,LL-73,OCKO-90}, in agreement with 
standard LGW arguments.

The behavior of the model (\ref{Hkin_kappa0})  for $p/r\not=2$ is less
clear.  Since gauge fields are not dynamical for $\kappa = 0$---they can be
integrated out---the critical behavior is completely determined by an 
effective model in terms of a gauge-invariant order parameter. 
In this effective description only the global symmetry group is relevant.
The gauge invariance group has the only role of selecting the appropriate
gauge-invariant order parameter. 
In the present case 
the  global symmetry group of the model is U($N_f$)/${\mathbb Z}_p$,
which is not a simple group. Thus, it is possible to have 
transitions where only the SU($N_f$) symmetry is broken, transitions where only
the Abelian subgroup U(1)/${\mathbb Z}_p$ is broken, and 
transitions where both subgroups condense simultaneously. Thus, one 
or two transitions may occur as a function of $\beta$. In the following 
we will discuss the behavior of the model for $Q=1,2$ and two values of 
$N_f$, $N_f=2$ and $N_f=15$. In all cases numerical results show that 
two transitions occur: a first one 
at $\beta_{c1}$ and a second one at $\beta_{c2}>\beta_{c1}$. 
At $\beta_{c1}$ the 
order parameter $T^{ab}$ condenses, while $\Sigma_q$ is still disordered: 
the transition is clearly associated with the breaking of the SU($N_f$) 
symmetry. At $\beta_{c2}$, $\Sigma_q$ correlations become critical,
signalling the breaking of the Abelian symmetry. The small-$\beta$ transition
has the same symmetry-breaking pattern of the small-$\kappa$ transitions 
observed 
in the model with U(1) gauge symmetry \cite{PV-19-AH3d}. Under the 
assumption that the nature of the transition only depends on the 
global symmetry breaking pattern, we can predict that the critical behavior 
is the same as that observed in 
Ref.~\cite{PV-19-AH3d}. For $N_f\ge 3$, only first-order
transitions are possible, under the usual assumption that a 
cubic term in the Landau-Ginzburg-Wilson (LGW)
 Hamiltonian signals discontinuous 
transitions (for a discussion of the validity of this assumption 
see Ref.~\cite{PV-19-AH3d} and references therein). For $N_f=2$, 
continuous transitions are possible: they belong to the O(3) vector
universality class.  The second, large-$\beta$ 
transition is associated with the 
breaking of the Abelian U(1) subgroup. Again, assuming that gauge fields are 
not dynamical, we predict that the transition belongs to the O(2)/XY
universality class, if it is continuous.

Let us finally consider the limit $\beta\to \infty$, In this case the 
scalar field orders so that 
\begin{equation}
{\bm w}_{\bm x} = \sigma_{{\bm x}+\hat{\mu}}^Q {\bm w}_{{\bm x} + \hat{\mu}},
\end{equation}
on each link. This relation implies 
\begin{equation}
{\bm w}_{\bm x} = \Pi_{{\bm x},\mu\nu}^Q {\bm w}_{\bm x}  \qquad
\Rightarrow  \qquad  \Pi_{{\bm x},\mu\nu}^Q = 1
\end{equation}
on all plaquettes, which, in turn, implies that $\Pi_{{\bm x},\mu\nu}$ 
belongs to the ${\mathbb Z}_M$ subgroup of ${\mathbb Z}_q$ ($M$ is 
the greatest common divisor of $Q$ and $q$). Thus, 
modulo gauge transformations, for $\beta\to \infty$ we can take 
$\sigma_{{\bm x},\mu} = \exp (2 \pi i n/M)$, with $n=0,\ldots M$. 
If $M=1$, the gauge fields are completely ordered on the line 
$\beta = \infty$. If, instead, $M\ge 2$, there is still a nontrivial 
gauge dynamics and the system behaves as a ${\mathbb Z}_M$ gauge model, which 
undergoes a finite-$\kappa$ topological phase transition.

\section{Numerical results: Critical behavior for $N_f = 2$ and $Q = 1$}
\label{sec5}

\begin{figure}
\begin{tabular}{lcr}
\includegraphics*[width=7cm,angle=0]{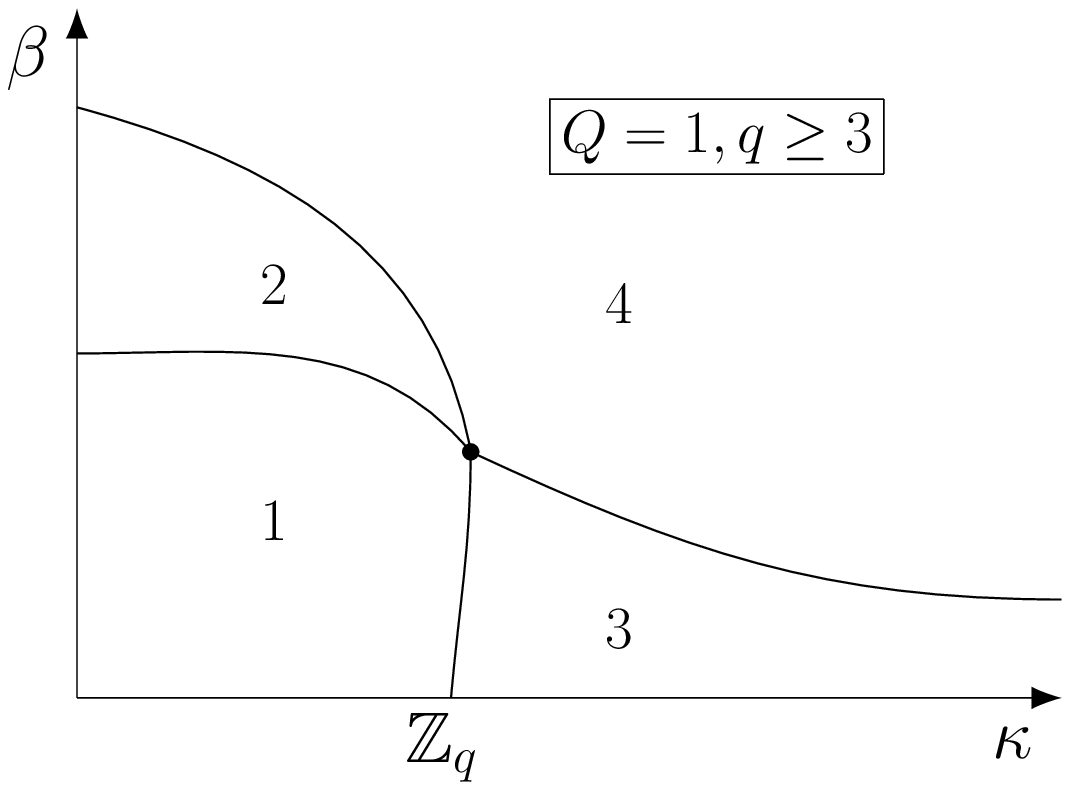} & $\qquad$ &
\includegraphics*[width=7cm,angle=0]{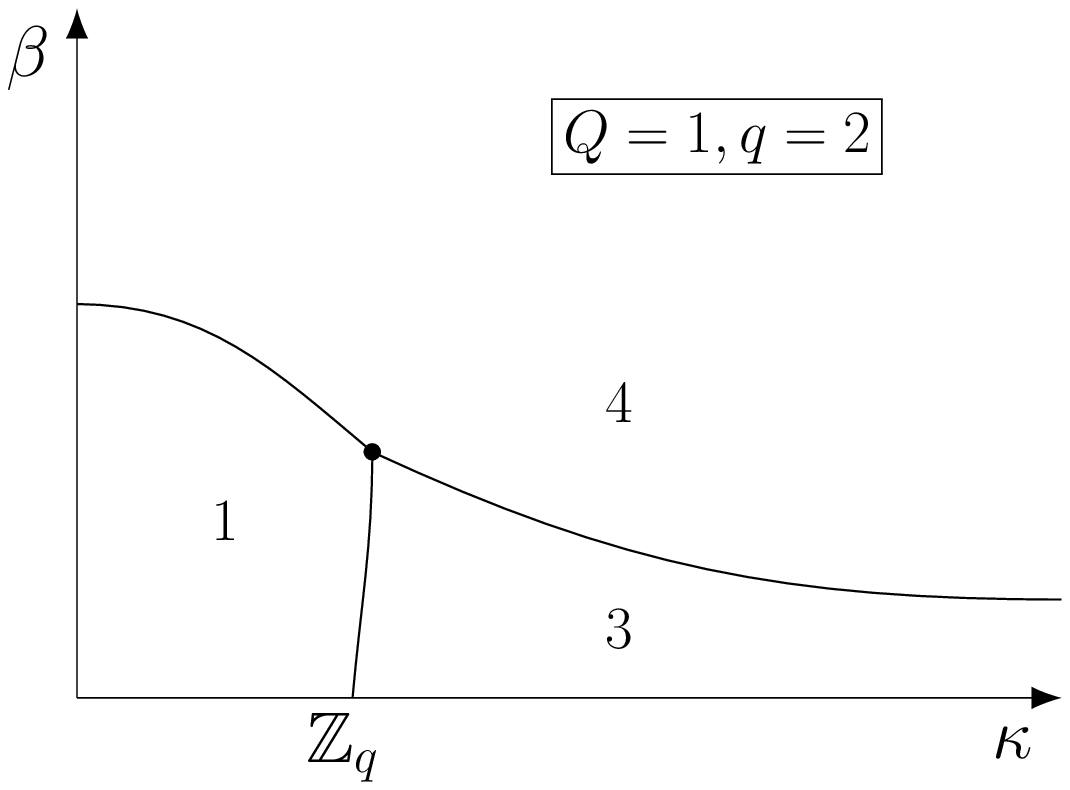}
\end{tabular}
\caption{Expected phase diagram of the model for $Q =1$. Left: $q\ge 3$; 
Right: $q = 2$. For $q \ge 3$ (left) there are four different phases. 
Scalar fields are disordered in phases 1 and 3, while 
they are ordered in phases 2 and 4. Transition lines 1-2 and 2-4 are
associated with the breaking of 
the global symmetry SU($N_f$) and of 
the global symmetry U(1)/${\mathbb Z}_q$, respectively; along the 
line 3-4 the simultaneous breaking of both  global symmetries occurs. 
A topological ${\mathbb Z}_q$ gauge transition occurs along the line 1-3. 
If $q = 2$ (right), 
phase 2 is missing and there is a single low-$\kappa$ transition line 1-4
along which the global vector O($2N_f$) symmetry is broken.
}
\label{phdiaQ1}
\end{figure}

In this section we discuss the phase diagram of the two-component
model ($N_f=2$) with charge-one ($Q=1$) scalar fields. We consider
three values of $q$, $q=2$, 3, and 4. The 
results that we shall discuss below are consistent with the phase diagrams
reported in Fig.~\ref{phdiaQ1}. 
As expected---the global symmetry group is different in the two cases---the 
phase diagram for $q=2$ is different from that of systems with $q\ge 3$. 

To identify the different transition lines,
we have performed runs at fixed $\kappa$, varying $\beta$. 
To determine the nature of the transitions that separate phases 1 and 2,
and phases 2 and 4 (see Fig.~\ref{phdiaQ1})
we have fixed $\kappa = 0.4$ for all $q$ values.
This value of $\kappa$ has been chosen on
the basis of the estimates of the critical $\kappa_c$ for the 
pure gauge model ($\beta = 0$). Indeed, we have 
$\kappa_c = 0.761413292(12)$ (using duality and the results 
of Ref.~\cite{FXL-18}), 1.0844(2), 1.52276(4) \cite{BCCGPS-14} for 
$q=2,$ 3, and 4, respectively. The value of $\kappa$ we use is significantly
smaller than the values of $\kappa_c$ reported above, guaranteeing
the we are indeed studying the small-$\kappa$ transition lines.
To investigate the large-$\kappa$ behavior we have instead fixed 
$\kappa = 1.5$ and 2 for $q = 2$ and 3, respectively.

In the runs at fixed $\kappa = 0.4$, we consider cubic lattices of
linear size $L$ with periodic boundary conditions. On the other hand, 
to determine the large-$\kappa$ critical behavior,
we consider open boundary conditions,
to avoid slowly-decaying dynamic modes that are present in systems with
periodic boundary conditions. Indeed, in the latter case, the 
Polyakov loops (the product of the gauge compact fields along nontrivial
lattice paths that wrap around the lattice) have a very slow dynamics,
if one uses algorithms with 
local updates. For open boundary conditions, Polyakov loops are not gauge 
invariant and thus their dynamics is not relevant for the estimation of 
gauge-invariant observables. A local algorithm is therefore efficient. 
Of course, open boundary conditions give rise to additional scaling
corrections, due to the boundary, and thus larger systems are needed 
to obtain asymptotic results.

\subsection{Small-$\kappa$ transitions}

\begin{figure}
\includegraphics*[width=5.5cm,angle=-90]{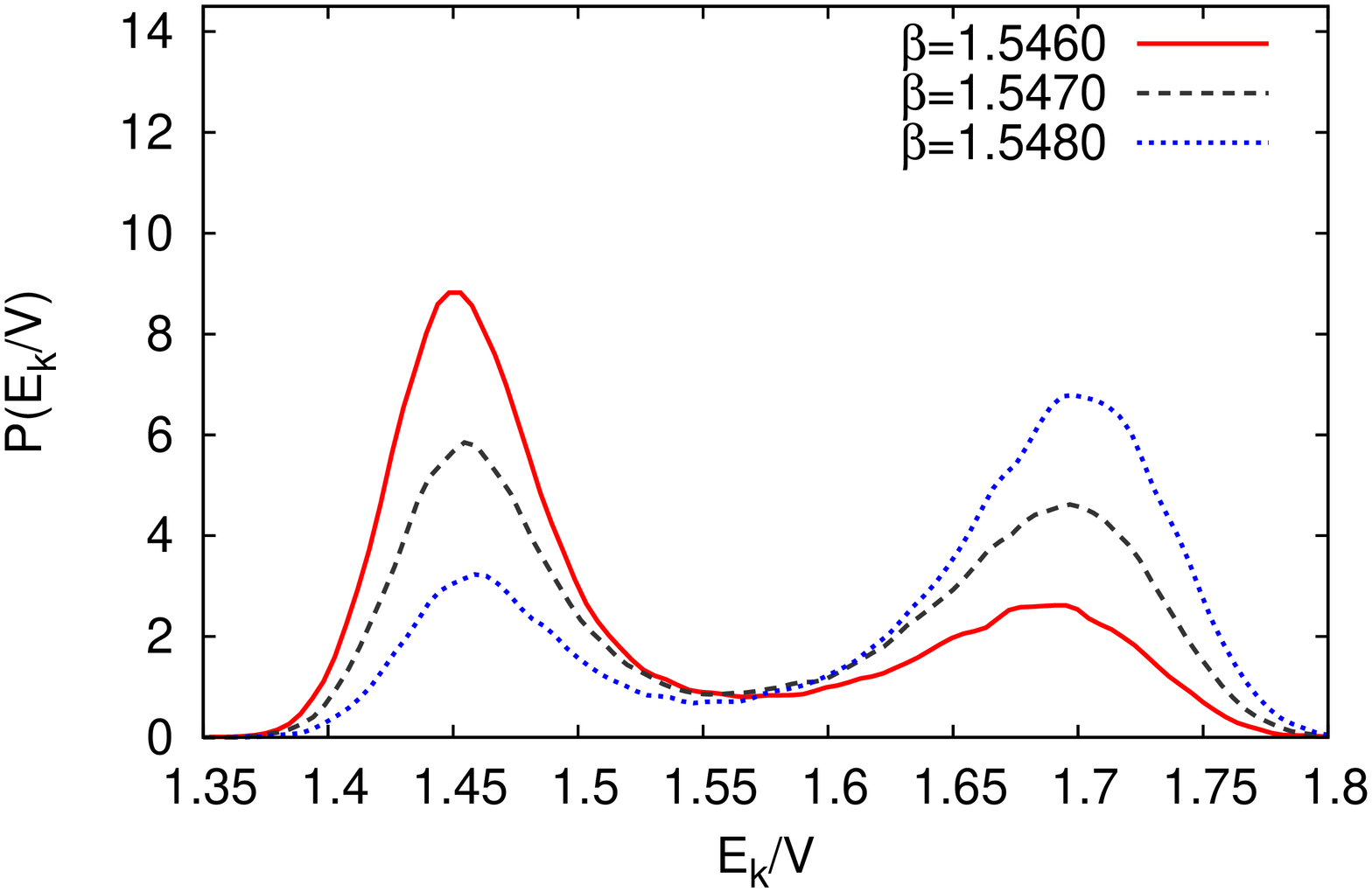}
\includegraphics*[width=5.5cm,angle=-90]{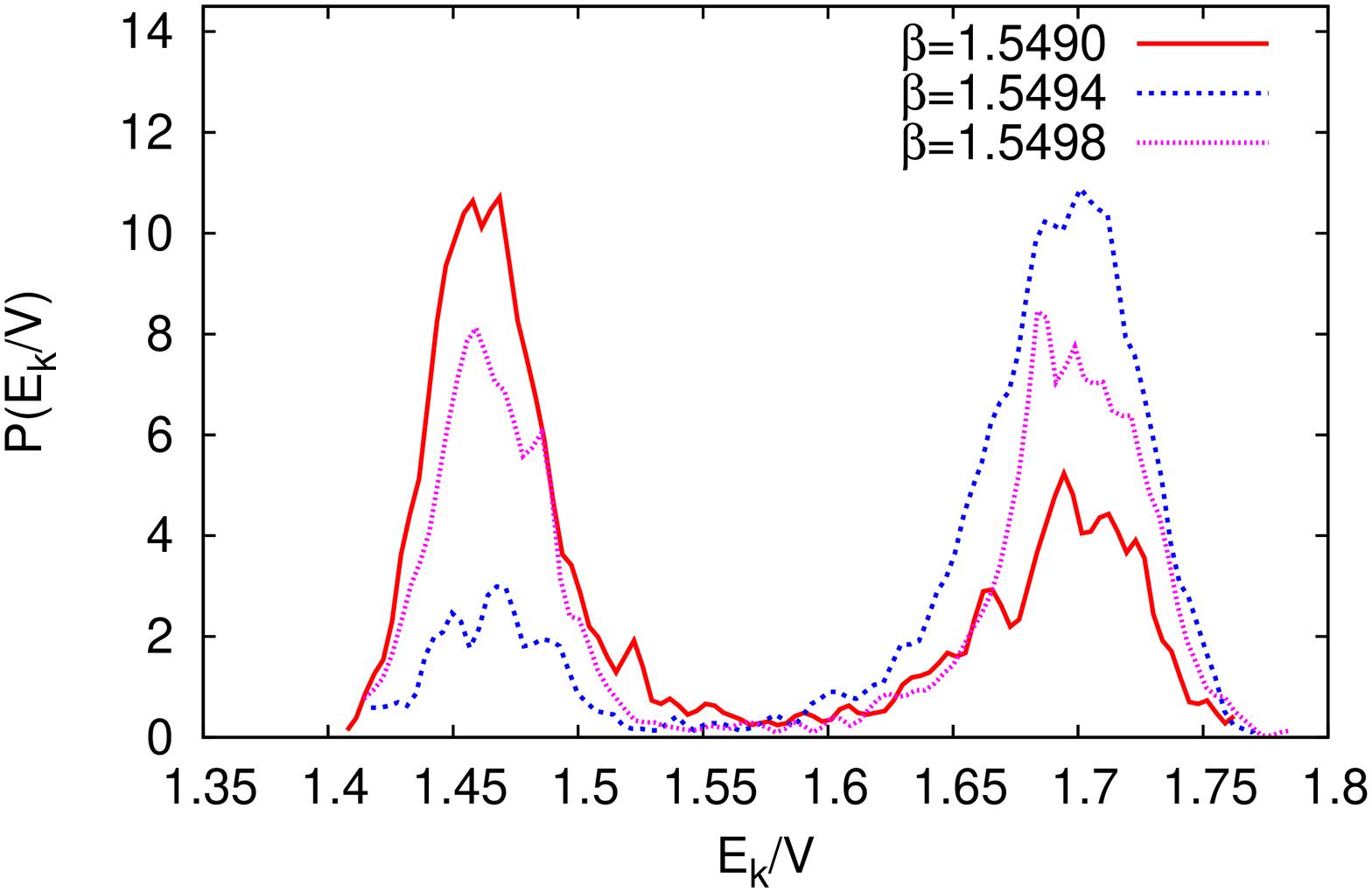}
\caption{Histograms of the scalar energy density $E_{k}/V$ for the 
model with $q = 2$, $N_f = 2$ along the line $\kappa=0.4$. 
Left: results for $L=16$; right: results for $L=20$. The $\beta$ values are 
reported in the legend.
}
\label{q2N2-lowk}
\end{figure}

\begin{figure}
\includegraphics*[width=5.5cm,angle=-90]{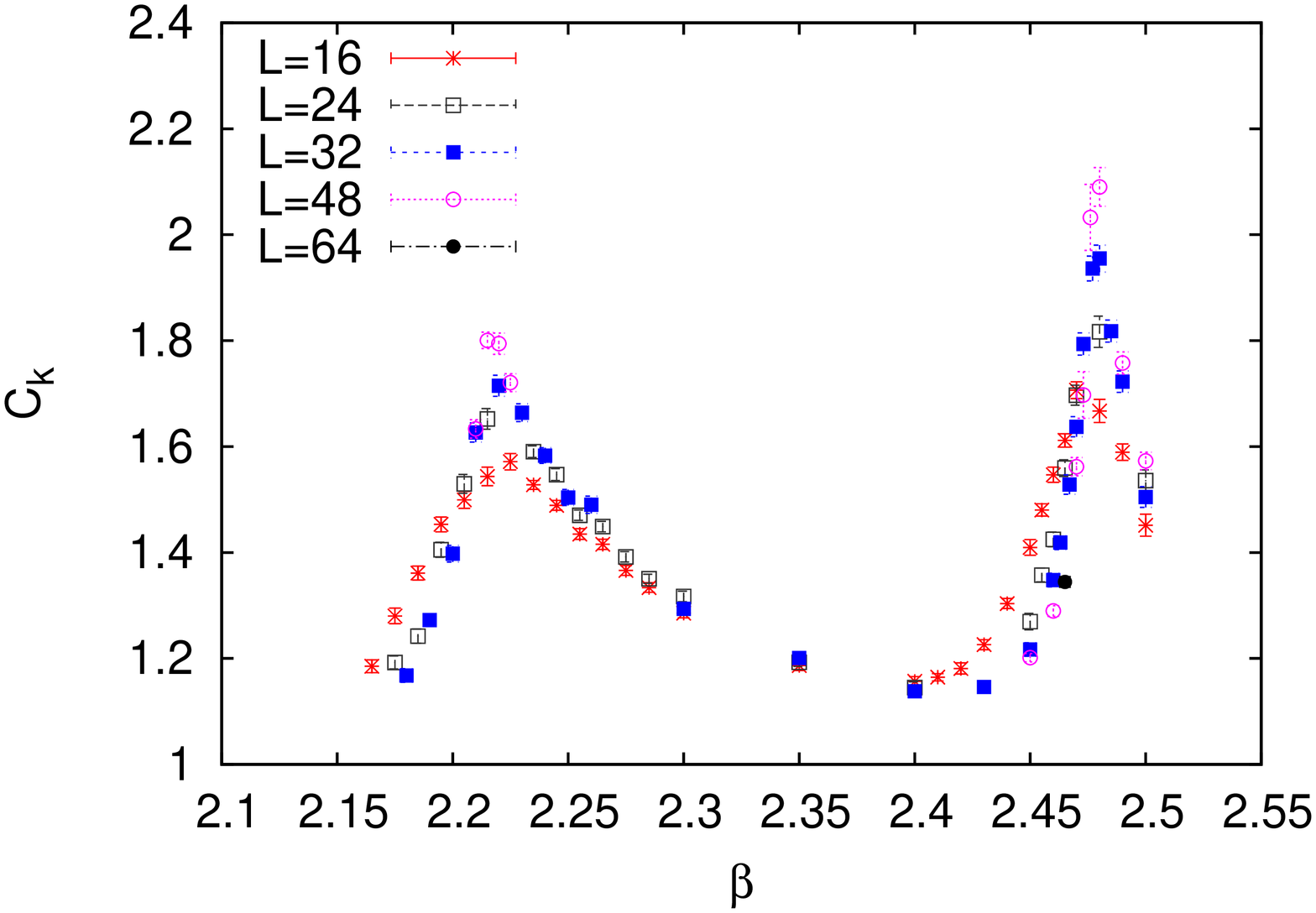}
\includegraphics*[width=5.5cm,angle=-90]{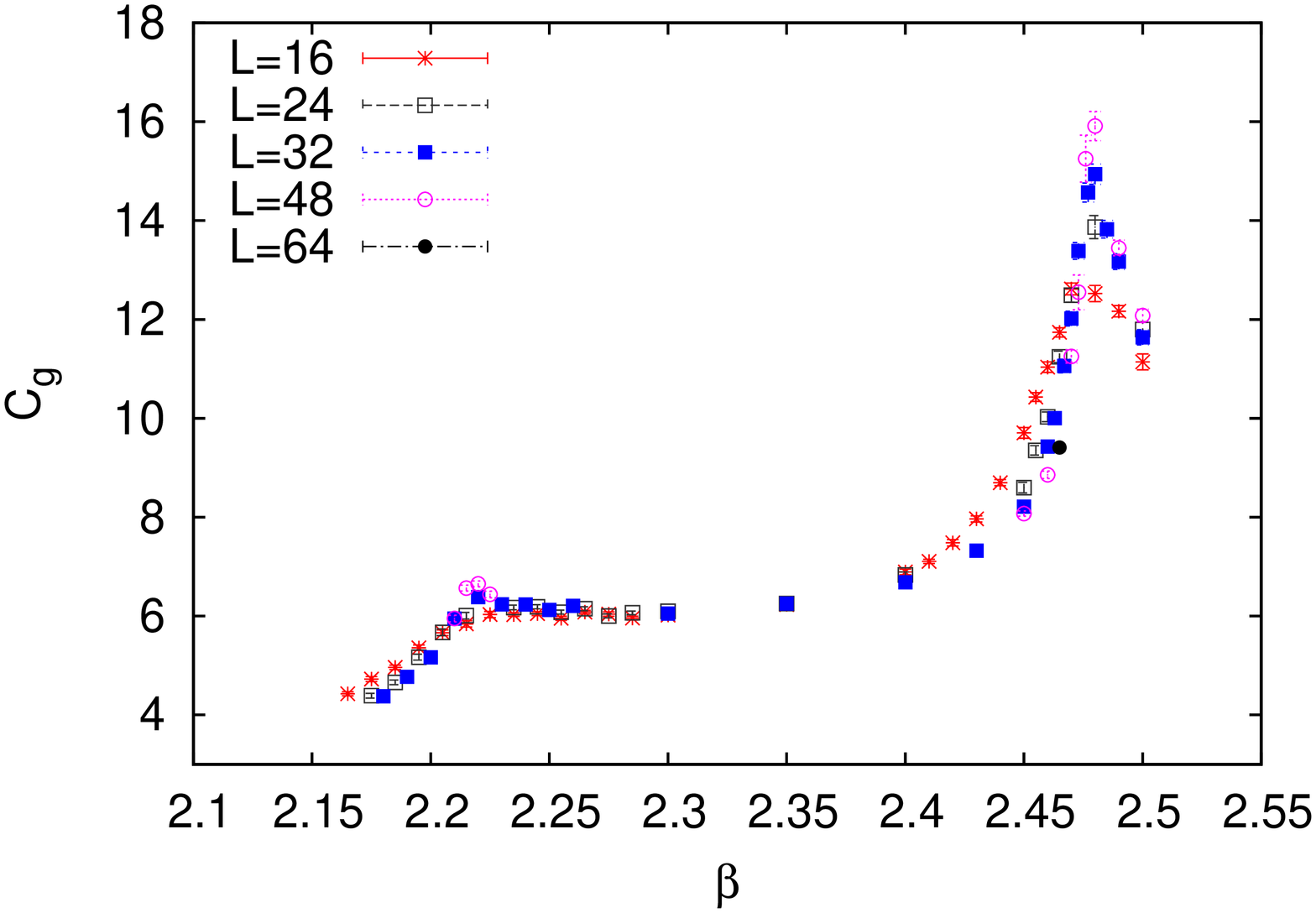}
\caption{
Plot of $C_k$ (left) and $C_g$ (right) versus $\beta$
for the two-flavor model ($N_f = 2$) with $q=3$ and $Q=1$, 
along the line $\kappa=0.4$.
}
\label{C-q3N2-lowk}
\end{figure}

\begin{figure}
\includegraphics*[width=5.5cm,angle=-90]{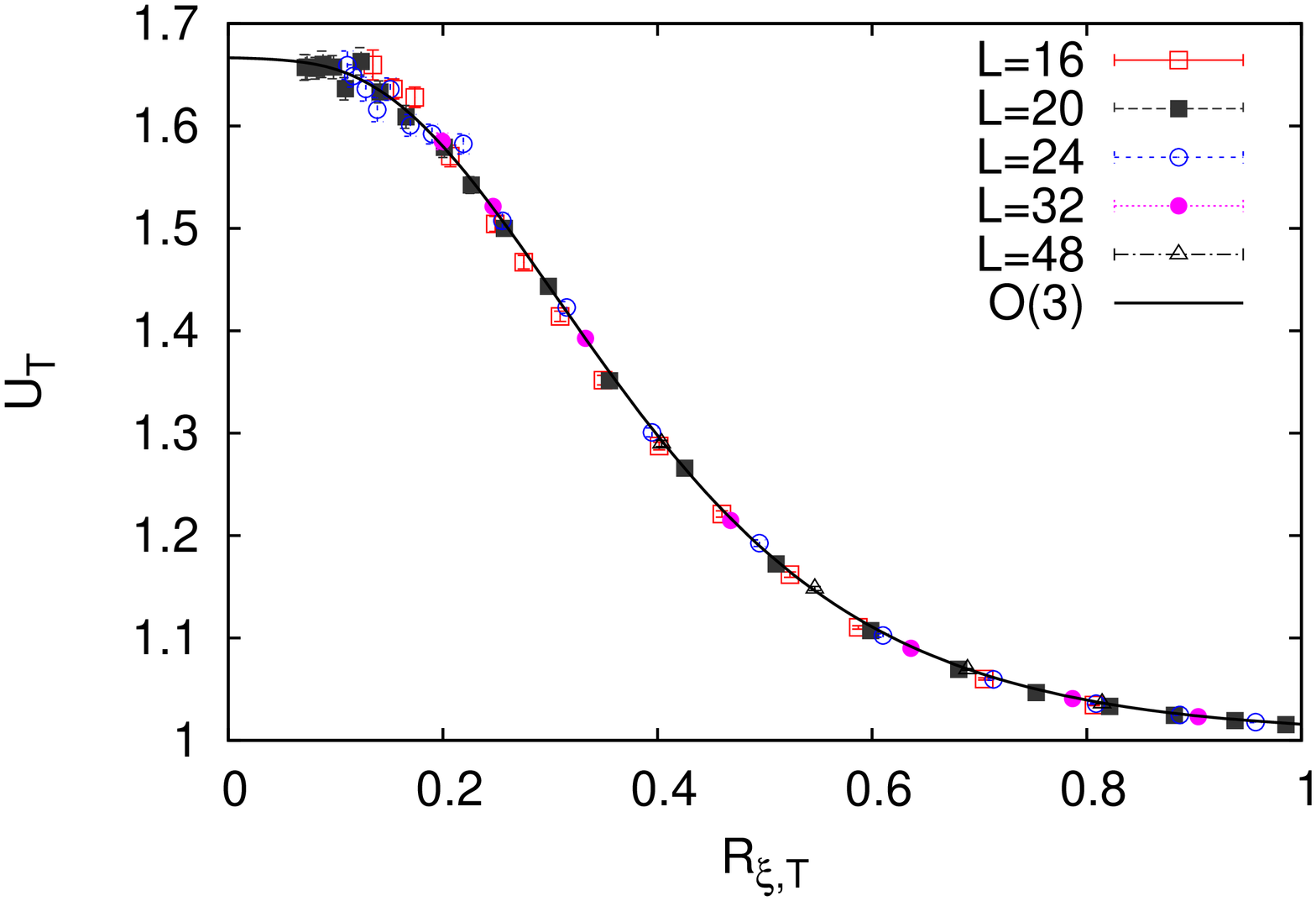}
\includegraphics*[width=5.5cm,angle=-90]{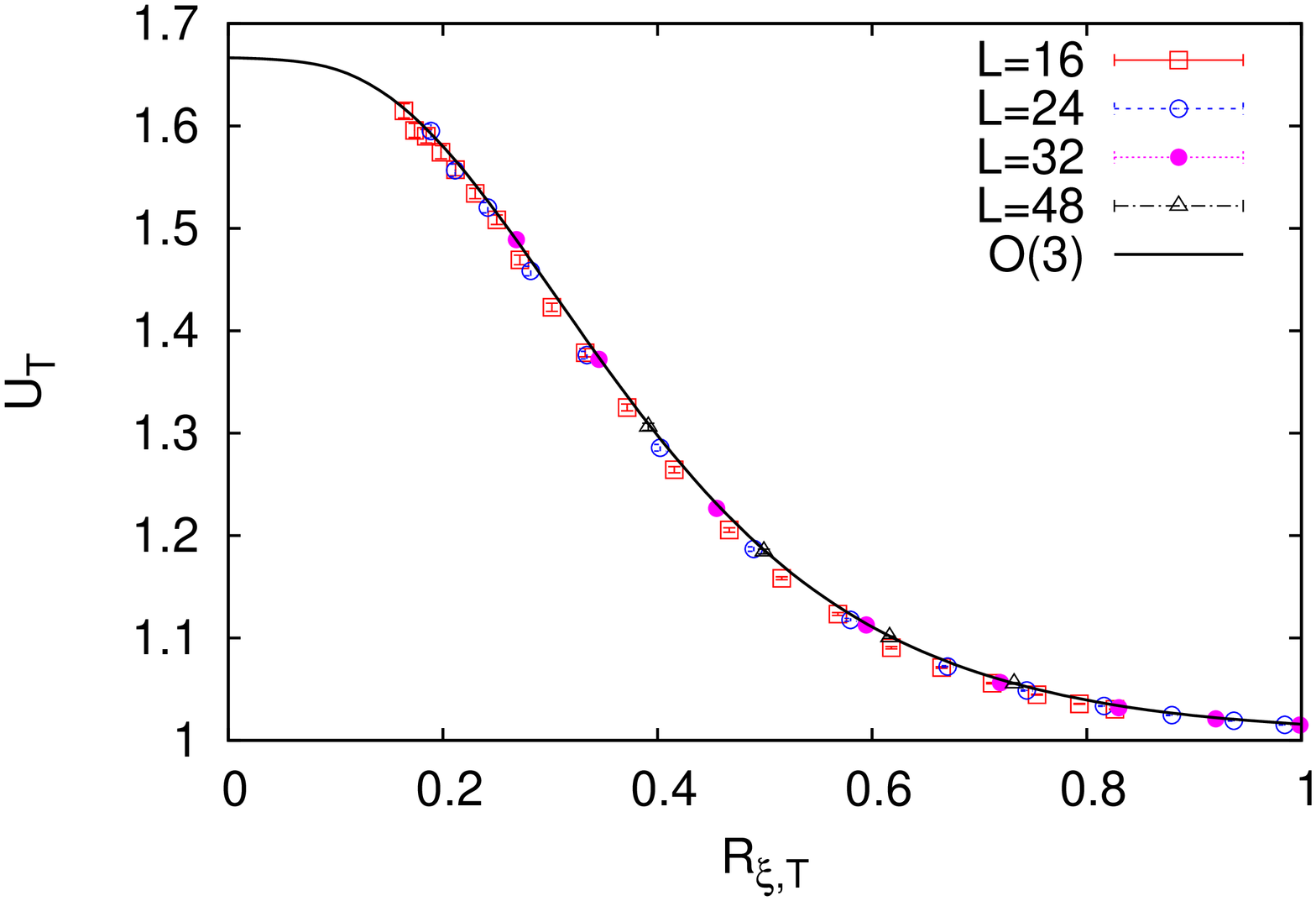}
\caption{
Plot of $U_T$ versus $R_{\xi,T}$ for the two-flavor model ($N_f = 2$) 
along the line $\kappa=0.4$ (small-$\beta$ transition): 
(left) results for $q=3$ (we only include 
data with $\beta < 2.3$); (right) results for 
$q=4$. The continuous line is the universal curve for vector 
correlations in the O(3) vector model 
(see Appendix of Ref.~\cite{BPV-21-gaugebr}).
}
\label{UvsRxi-q34N2-lowk}
\end{figure}

\begin{figure}
\begin{center}
\includegraphics*[width=6cm,angle=-90]{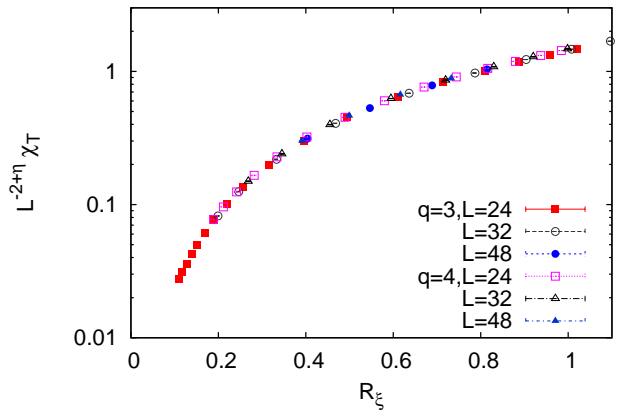}
\end{center}
\caption{
Plot of $L^{-2+\eta}\chi_T$ versus $R_{\xi,T}$ for 
the two-flavor model ($N_f = 2$) 
along the line $\kappa=0.4$ (small-$\beta$ transition): 
results for $q=3$ (we only include 
data with $\beta < 2.3$) and $q=4$. We use
the O(3) value~\cite{Hasenbusch-20} $\eta = 0.03624$.
}
\label{chiT-q34N2-lowk}
\end{figure}

We have first considered the $q=2$ model, which has an
enlarged O(4) global symmetry. Along the line $\kappa = 0.4$, 
simulations show the presence 
of a single first-order transition. The discontinuous nature is 
apparent from 
plots of the data as a function of the Monte Carlo
time, which show  the typical see-saw behavior, and from 
the histograms of the energy, which  have a clear bimodal structure,
see Fig.~\ref{q2N2-lowk}.
This result is in agreement with the general theory.
Indeed, for $q=2$ the model is equivalent to an 
RP$^3$ model, which is expected to undergo a first-order transition
\cite{dGP-93,LL-73,OCKO-90}. 

Let us now  discuss the behavior for $q\ge 3$.
As we have discussed in Sec.~\ref{sec4}, for $\kappa = 0$, and therefore 
also for small values of $\kappa$, it is possible to have two different 
transitions as a function of $\beta$, associated with different breakings of
the global symmetry. To verify this possibility, we have performed 
a scan in $\beta$. In Fig.~\ref{C-q3N2-lowk}, we report the specific heats 
as a function of $\beta$ for $q=3$. 
There is a clear evidence of two transitions,
one at $\beta \approx 2.2$, weakly coupled with the gauge degrees of 
freedom---the gauge specific heat is almost constant in the transition
region--- and a second one at $\beta \approx 2.45$. For $q=4$, we again
observe two transitions, at $\beta \approx 2.3$ and 4.35, respectively. 

We first focus on the transition at $\beta \approx 2.2$ for $q=3$. We find that,
while tensor correlations are critical ($\xi_T$ increases rapidly
at the transition), the gauge-invariant charge-3 correlations are always
short-ranged. This clearly indicates that the transition is associated
with the breaking of the SU(2) global symmetry.
In Fig.~\ref{UvsRxi-q34N2-lowk} we report $U_T$ versus $R_{\xi,T}$ 
and compare the results with the universal curve appropriate for the 
O(3) universality class, reported in the Appendix of Ref.~\cite{BPV-21-gaugebr}.
Data fall quite precisely onto the O(3) curve, confirming the LGW 
prediction that the transition belongs to the 
O(3) universality class. Also for $q=4$ 
the data close to the small-$\beta$ transition fall 
on top of the O(3) curve, indicating that the transition belongs to the 
O(3) universality class for any $q>3$, a result which is not surprising 
as the same O(3) transition occurs in the model with U(1) symmetry, i.e.,
for $q\to \infty$.

As an additional check we have analyzed
$R_{\xi,T}$ and $U_T$, fitting the Monte Carlo data in the small-$\beta$ 
transition region ($\beta < 2.3$ for $q=3$) 
to Eq.~(\ref{rsca}).  Scaling corrections have been neglected---as apparent from
Fig.~\ref{UvsRxi-q34N2-lowk}, they are small. The scaling function
$F_R(x)$ has been parametrized with a polynomial. We obtain
$\nu = 0.716(13)$ and 0.72(2) for $q=3,4$, respectively, 
in good agreement with the O(3) 
estimate $\nu = 0.71164(10)$ of Ref.~\cite{Hasenbusch-20}. To estimate the 
critical coupling $\beta_c$ we have repeated the same fits fixing 
$\nu$ to the O(3) value. We obtain
$\beta_c = 2.2155(3)$ and 2.3175(5) for $q=3,4$. Note that $\beta_c$ 
depends weakly on $q$, indicating that the nonuniversal features of the 
transitions are only slightly dependent on $q$.

Finally, we have checked the behavior of the tensor susceptibility 
that is expected to scale as $L^{2-\eta} f(R_{\xi,T})$, 
where $f(x)$ is universal
apart from a multiplicative rescaling. If the transition belongs to 
the O(3) universality class, data should scale provided we set
$\eta=\eta_{O(3)}$, where $\eta_{O(3)}  = 0.03624(8)$
\cite{Hasenbusch-20} is the value it takes in the O(3) model. 
Results are shown in 
Fig.~\ref{chiT-q34N2-lowk}. We observe an excellent scaling. Moreover, on the 
scale of the figure, we do not see any dependence of the scaling curve on $q$.
As observed for $\beta_c$, the $q$-dependence of nonuniversal amplitudes
is small.

\begin{figure}
\includegraphics*[width=5.5cm,angle=-90]{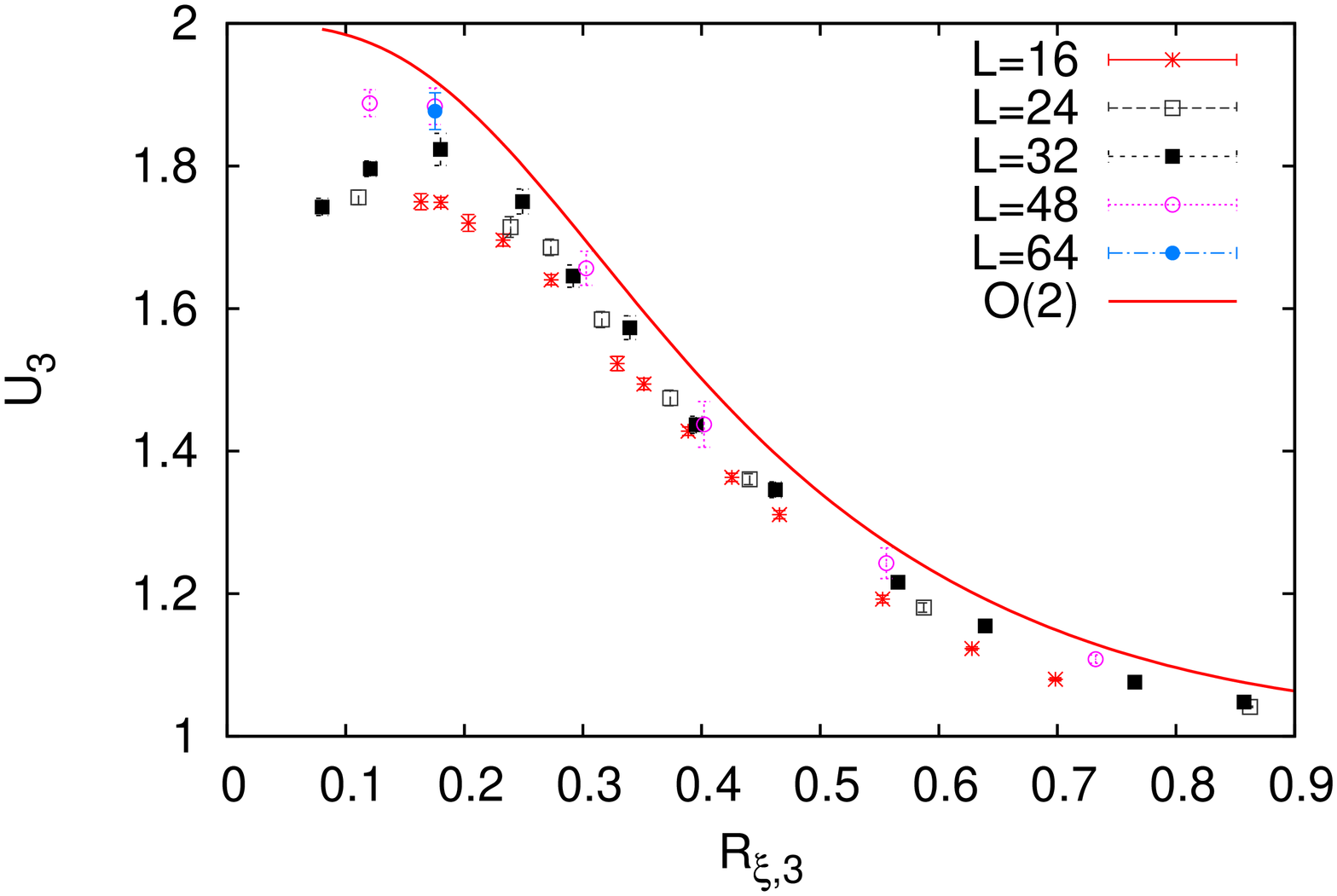}
\includegraphics*[width=5.5cm,angle=-90]{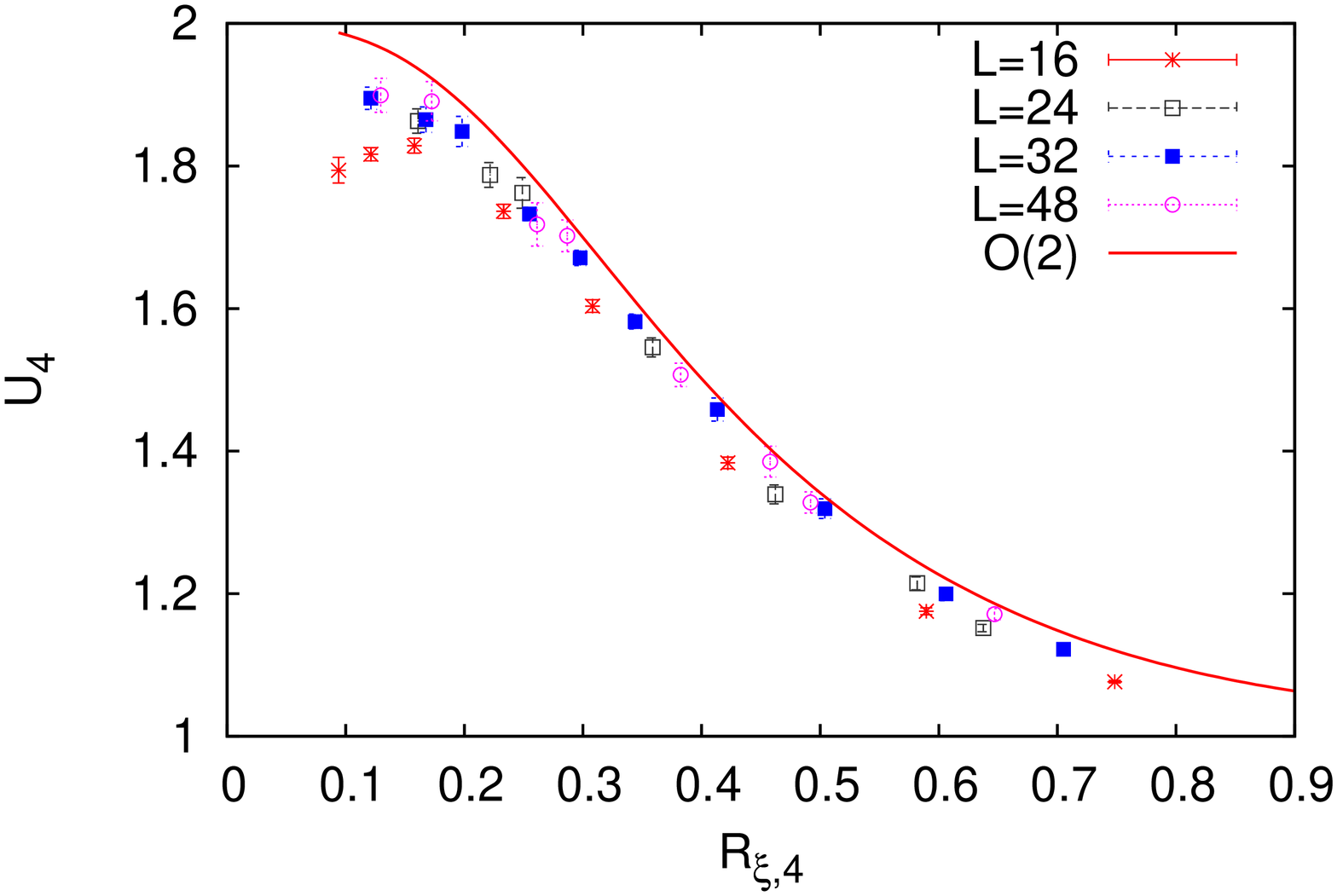}
\caption{Scaling plots at the large-$\beta$ critical transition, at fixed 
$\kappa = 0.4$, for $N_f=2$ and $Q=1$.
Left: plot of $U_3$ versus $R_{\xi,3}$ for $q=3$;
right: plot of $U_4$ versus $R_{\xi,4}$ for $q=4$.
We also report the universal scaling curve for vector correlations 
in the O(2) vector model (see Appendix of Ref.~\cite{BPV-21-gaugebr}).
}
\label{U1transition-N2-smallk}
\end{figure}

\begin{figure}
\includegraphics*[width=5.5cm,angle=-90]{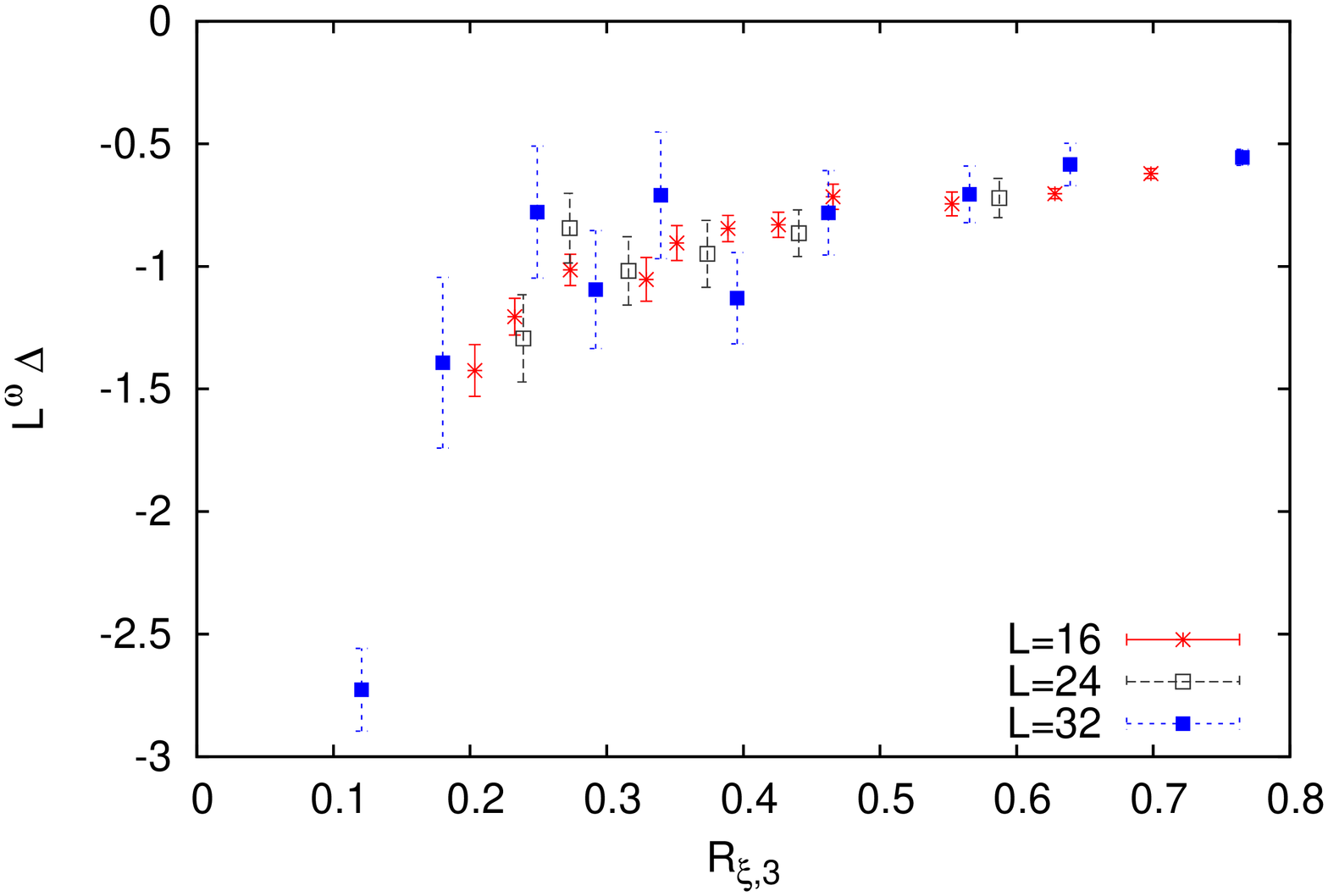}
\includegraphics*[width=5.5cm,angle=-90]{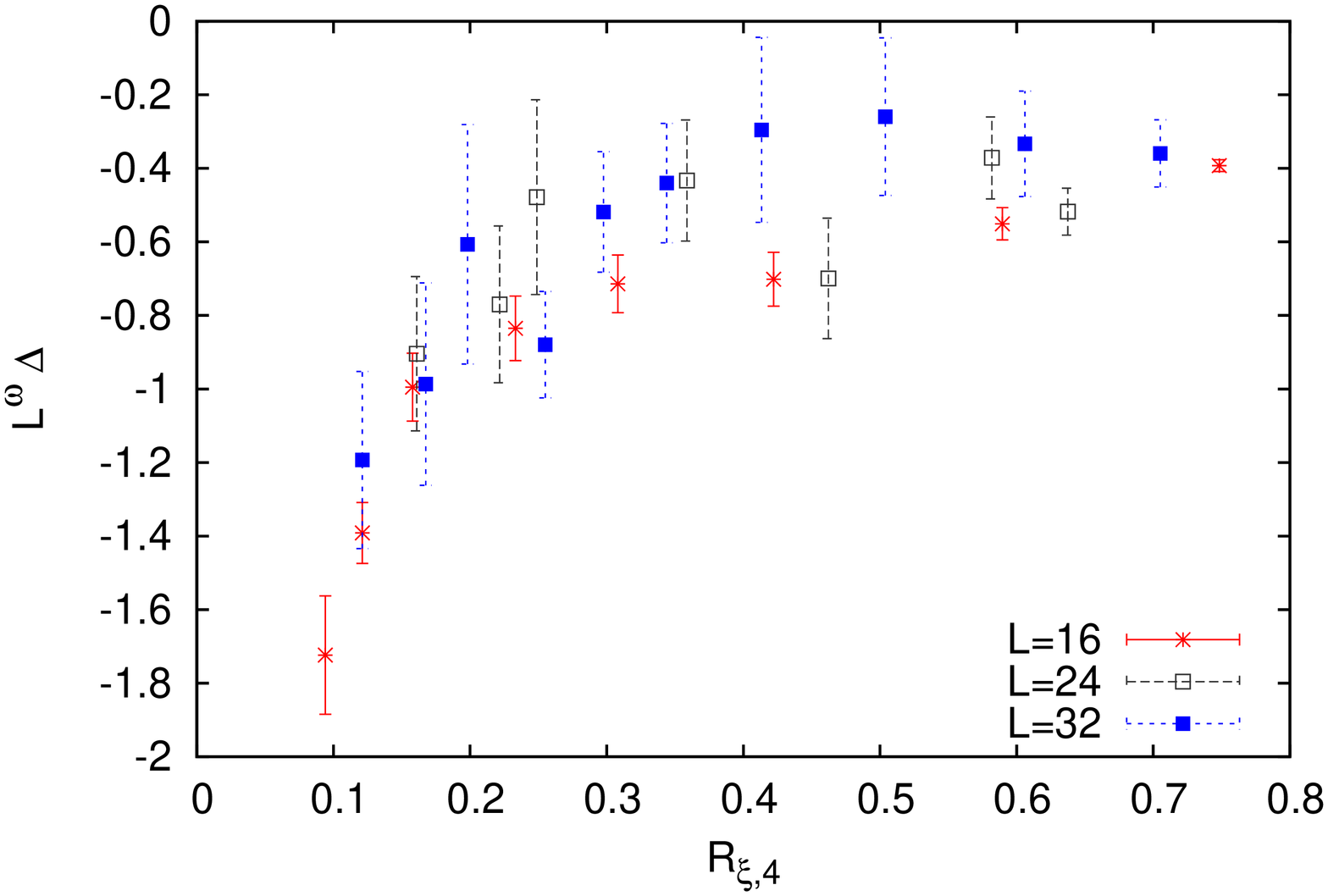}
\caption{Scaling corrections at the large-$\beta$ critical transition, at fixed 
$\kappa = 0.4$, for $N_f=2$ and $Q=1$.
Plot of $L^{\omega} \Delta(R_{\xi,q})$ versus $R_{\xi,q}$ for $q=3$ (left)
and $q=4$ (right), where $\Delta(R_{\xi,q})$ is defined in
Eq.~(\ref{Delta-def}). 
We set $\omega=\omega_{XY}$, where $\omega_{XY}$ is 
the XY value of the correction-to-scaling
exponent: $\omega_{XY} = 0.789$ \cite{Hasenbusch-19}.
}
\label{U1transition-dev-N2-smallk}
\end{figure}

Let us now focus on the large-$\beta$ transition. For both $q=3$ and 4, 
charge-$q$ correlations diverge, Since the $SU(N_f)$ symmetry is broken, 
the transition is described by an effective scalar model in which the
fundamental field is a phase and the interaction is ${\mathbb Z}_q$ 
gauge invariant. The behavior of this model has been discussed in 
Ref.~\cite{BPV-22-discrete}. For small values of $\kappa$, the transition
belongs to the XY universality class, if it is continuous.
Moreover, the universal scaling function that expresses $U_q$ in terms 
of $R_{\xi,q}$ is the same as the scaling fuction $U=f_U(R_\xi)$ for 
vector correlations in the XY model. To verify this prediction,
we plot $U_q$ versus $R_{\xi,q}$ for $q=3,4$ and compare the data with 
the universal XY vector curve reported in Ref.~\cite{BPV-21-gaugebr}, 
see Fig.~\ref{U1transition-N2-smallk}. 
Data are consistent with the XY curve (as $L$ increases,
data move closer to the XY curve), although significant
corrections are apparently present, 
especially for $q=3$, probably because of the
presence of the nearby O(3) transition. In order to have a stronger check
of the correctness of the prediction, we have analyzed the scaling corrections.
We consider the deviations
\begin{equation}
\Delta(R_{\xi,q}) = U_q - f_{XY}(R_{\xi,q}), 
\label{Delta-def}
\end{equation}
as a function of $R_{\xi,q}$; here $f_{XY}(R_{\xi})$ is 
the universal XY curve for vector correlations. 
If $f_{XY}(R_{\xi,q})$ is the correct 
asymptotic behavior and 
$\omega$ is the leading correction-to-scaling exponent, 
$L^\omega \Delta(R_{\xi,q})$ should scale for large $L$. 
We assume that $\omega$ is the same as the leading exponent $\omega_{XY}$ 
in the XY model.\footnote{{\em A priori} it is possible that 
$\omega < \omega_{XY}$, 
since in our model there are additional RG subleading operators due, e.g., 
to the gauge interactions, the interactions with the frozen SU$(N_f)$ 
modes, etc.. They may be more relevant than the RG operator that 
controls the scaling corrections in the XY model . Our numerical 
results are in agreement with the assumption $\omega = \omega_{XY}$.} 
In Fig.~\ref{U1transition-dev-N2-smallk} we report the numerical results.
Data scale reasonably, confirming that the transition belongs to the 
XY universality class. 

Assuming an XY behavior, we have estimated the location of the critical 
transition. We have performed combined fits of $U_q$ and $R_{\xi,q}$ 
to Eq.~(\ref{rsca}), 
including scaling corrections. Fixing $\omega$ and $\nu$ to the XY values,
$\omega = 0.789$, $\nu = 0.6717$ \cite{Hasenbusch-19}, 
we obtain $\beta_c = 2.4769(4)$ and $\beta_c = 4.3370(15)$ for 
$q=3$ and 4, respectively.

\subsection{Large-$\kappa$ transition line}

\begin{figure}
\includegraphics*[width=5.5cm,angle=-90]{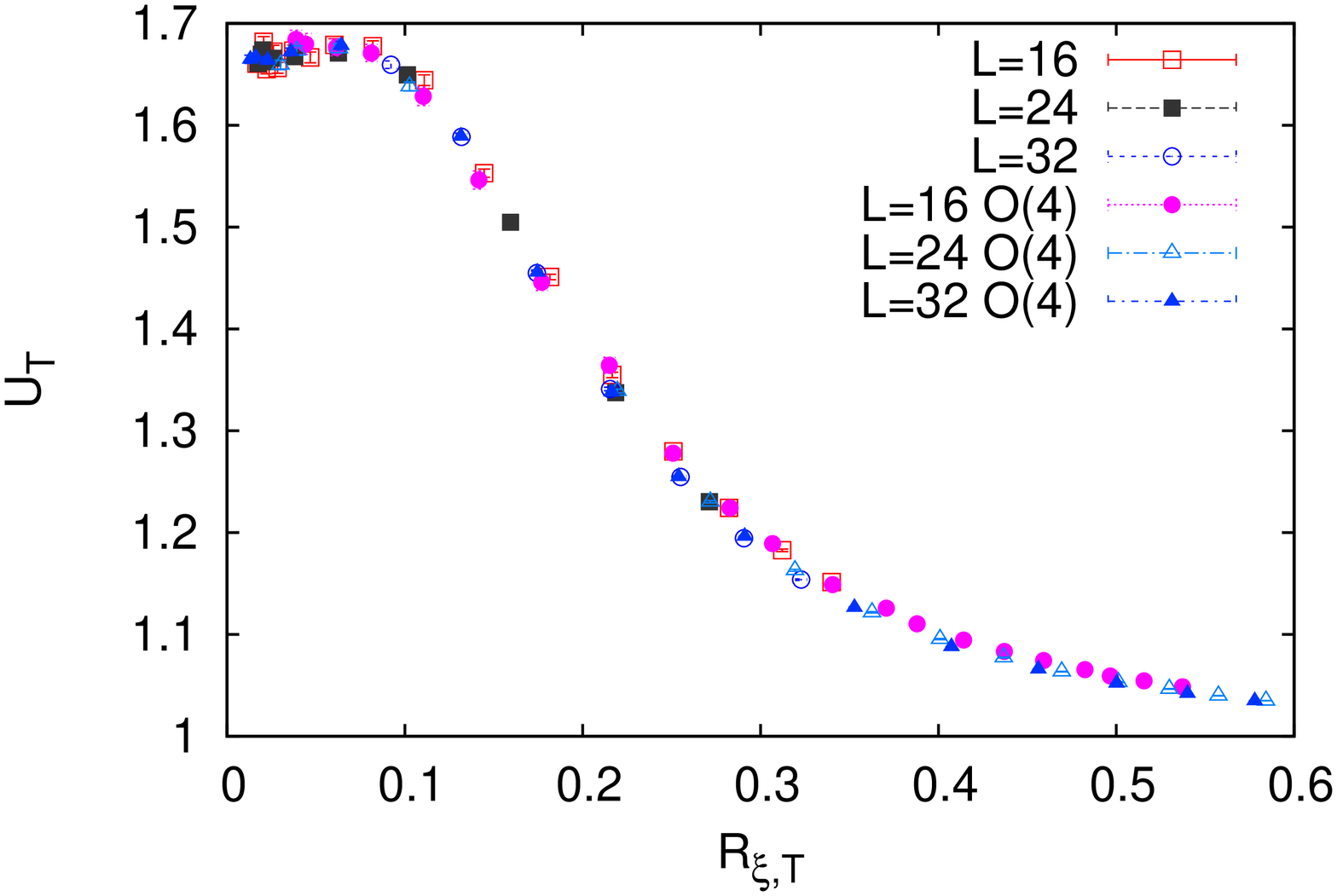}
\includegraphics*[width=5.5cm,angle=-90]{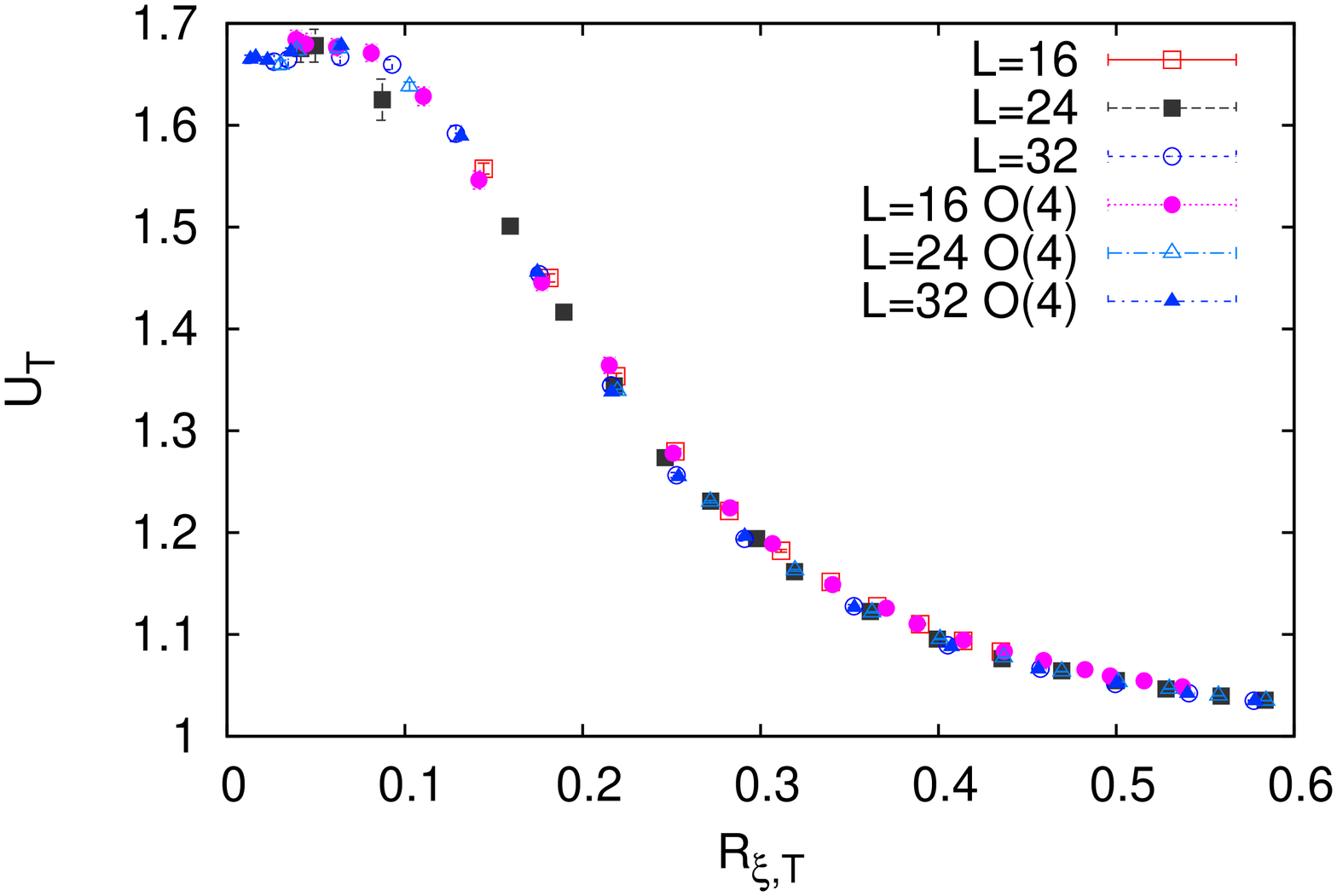}
\caption{
Plot of $U_T$ versus $R_{\xi,T}$ for the two-flavor model ($N_f = 2$) 
along the large-$\kappa$ transition line (open boundary conditions). 
Left: results for $q=2$ (at fixed 
$\kappa = 1.5$); right: results for $q=3$ (at fixed
$\kappa = 2.0$). The results for the gauge model are compared with the 
results for the O(4) vector model.
}
\label{RxiTUT-q23N2-largek}
\end{figure}

\begin{figure}
\begin{center}
\includegraphics*[width=6cm,angle=-90]{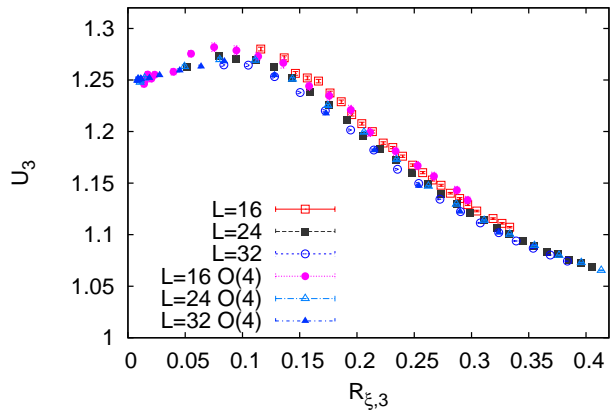}
\end{center}
\caption{
Plot of $U_3$ versus $R_{\xi,3}$ for the two-flavor model ($N_f = 2$) 
along the large-$\kappa$ transition line, for $q=3$ (at fixed
$\kappa = 2$). 
The results for the gauge model are compared with the 
results for the O(4) vector model.
}
\label{Rxi3U3-q3N2-largek}
\end{figure}

We have studied the large-$\kappa$ behavior of the system for $q=2$ and $q=3$, 
performing runs at fixed $\kappa = 1.5$ and $\kappa = 2$, respectively. 
As we discussed above, we expect the same behavior as in the 
O(4) vector model, for all gauge-invariant observables. Note that, for $q=3$
this represents an enlargement of the global symmetry at the transition.
In the presence of O(4) symmetry, 
tensor and charge-$q$ correlations are related, see Sec.~\ref{sec3},
and thus they should be both critical along the large-$\kappa$ 
transition line. 

The
numerical results are in agreement with these predictions. 
In Fig.~\ref{RxiTUT-q23N2-largek} we compare the behavior of the tensor
Binder parameter $U_T$ as a function of $R_{\xi,T}$ in the gauge models 
and in the O(4) model. Data scale onto the same universal curve, as expected.
We have also verified that charge-$q$ correlations behave as in the 
O(4) model. For $q=2$ this is a consequence of the exact O(4) symmetry of the 
model. For $q=3$ this is confirmed by the results shown  
in Fig.~\ref{Rxi3U3-q3N2-largek}.

\subsection{Phase diagram}

The numerical results presented above confirm that, for $N_f=2$, 
the model has the phase diagram reported 
in Fig.~\ref{phdiaQ1}. We expect the same phase diagram for 
any $N_f$ as long as $Q = 1$.
As far as the nature of the transition lines, for $q\ge 3$ we expect 
the small-$\kappa$, large-$\beta$ transition, 
where the U(1)/$\mathbf{Z}_q$ symmetry
is broken, to belong to the XY universality class in all cases. 
Analogously, the large-$\kappa$ transitions should always belong to the 
O($2N_f$) universality class. 
For small $\kappa$ and small $\beta$, transitions should 
be analogous to that observed in the CP$^{N_f-1}$ model, and therefore 
they should be of first order for any $N_f \ge 3$.

If $Q\not=1$ but $M$ is 1 ($M$ is the greatest common divisor of $Q$ and $q$),
the model should also have the phase diagram reported in Fig.~\ref{phdiaQ1}.
Instead, for $M \ge 2$ the phase diagram is more complex (it will be discussed
in the next section), because of the presence of a 
new topological transition line for large 
values of $\beta$.

\section{Numerical results: Critical behavior for $N_f = 15$ and $Q = 2$}
\label{sec6}

\begin{figure}
\begin{tabular}{lcr}
\includegraphics*[width=7cm,angle=0]{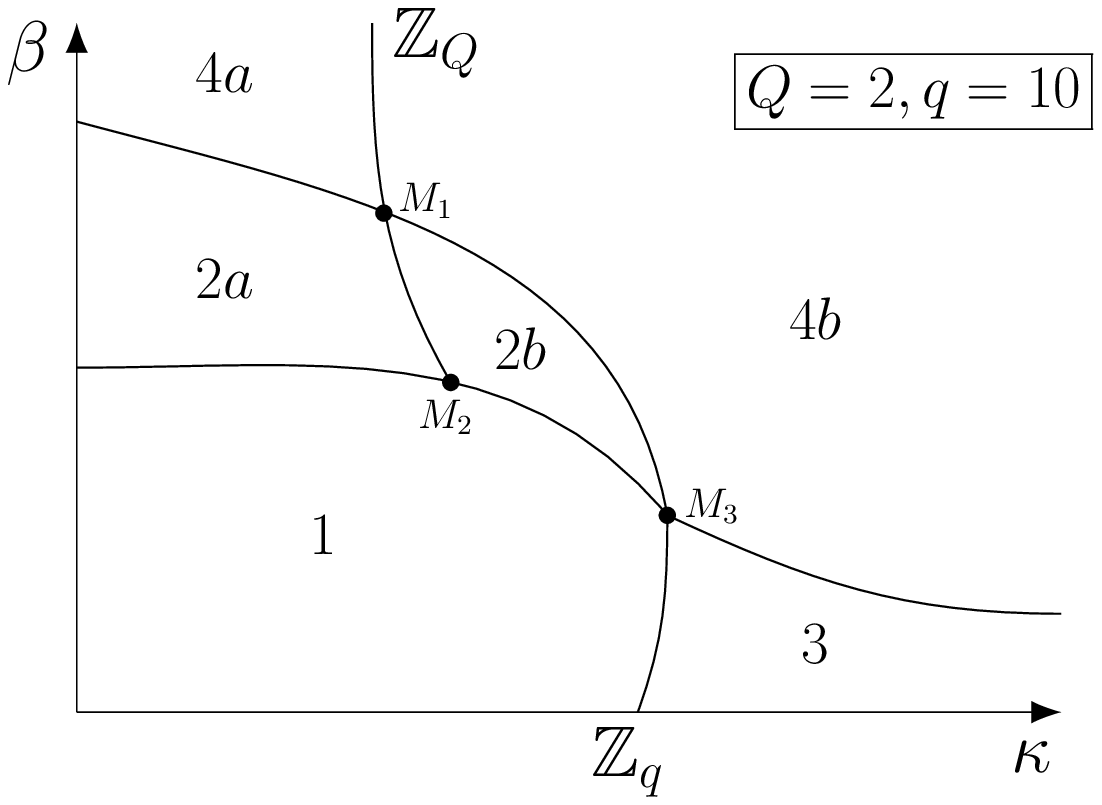} & $\qquad$ &
\includegraphics*[width=7cm,angle=0]{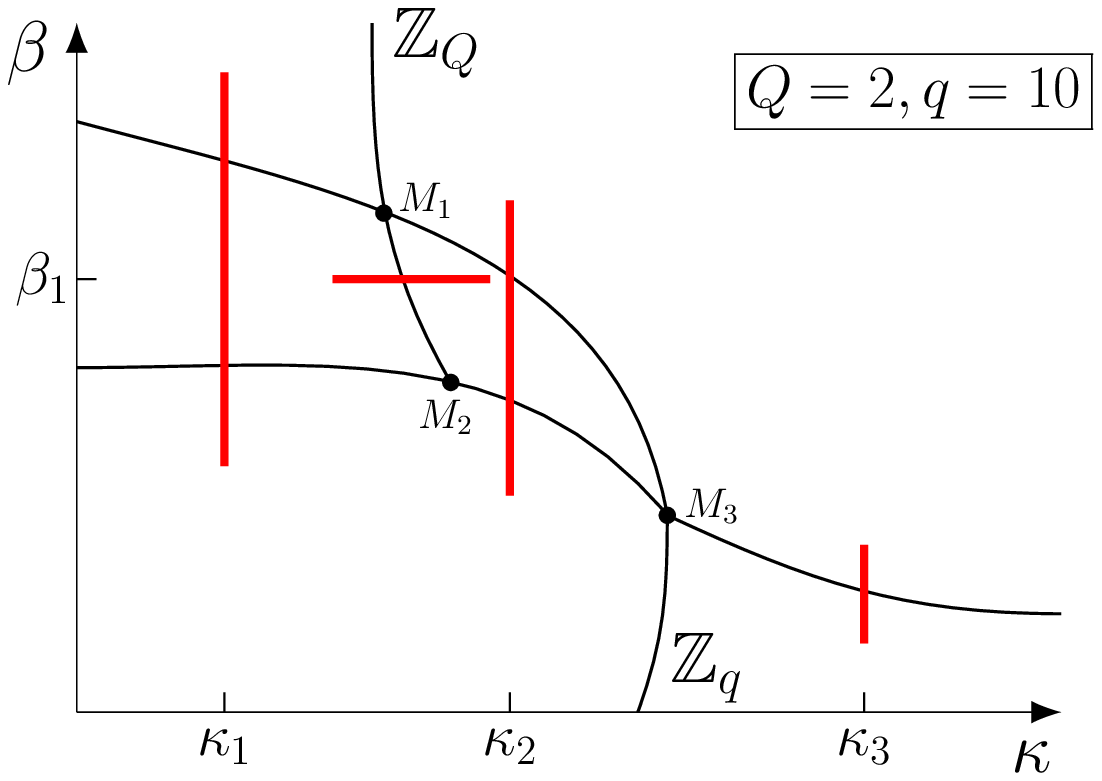}
\end{tabular}
\caption{Left: sketch of the phase diagram of the model for $Q =2$ and $q = 10$.
We expect the same phase diagram for any even $q\ge 6$. For $q=4$, the system
has a larger O($2N_f$) global symmetry and phases $2a$ and $2b$ are missing.
In the right panel we report
the lines (red thick lines) along which runs have been performed.
We have performed runs at fixed $\kappa$, at $\kappa_1 = 0.4$, $\kappa_2 =
1.2$, and $\kappa_3 = 10$, and a series of runs at fixed $\beta$, at
$\beta = \beta_1 = 13$. The tensor correlation length diverges in phases
$2a$, $2b$, $4a$, $4b$, signalling the breaking of the SU($N_f$) symmetry;
the correlation length $\xi_5$ is finite in phases $2a$, $2b$, and 
diverges in phases $4a$, $4b$.
}
\label{phdiaQ2}
\end{figure}

In this Section, we consider the model with charge $Q=2$ and
$N_f=15$ for different values of $q$ with the purpose of understanding whether
it exhibits a continuous transition in the same universality class 
as the charged transition that occurs in  U(1) gauge invariant models
\cite{BPV-21-NCQED,BPV-20-AHq2,BPV-22-AHq}. We only consider even values of
$q$. If $q$ is odd, gauge fields are completely ordered for $\beta \to \infty$.
In this case, the phase diagram should be analogous to that obtained for $Q=1$,
see Fig.~\ref{phdiaQ1}, without the large-$\beta$ topological transition line,
whose presence is necessary to observe the charged transition in 
U(1) gauge invariant models
\cite{BPV-21-NCQED,BPV-20-AHq2,BPV-22-AHq}. In most of the simulations
we set $q=10$. The numerical results are 
consistent with the phase diagram reported in 
Fig.~\ref{phdiaQ2}. Comparing with Fig.~\ref{phdiaQ1}, one observes the
presence of a new line that starts at $\beta = \infty$ and that 
is associated with 
a ${\mathbb Z}_Q$ (in our case Ising) topological transition. It
intersects the two phases labelled 2 and 4 in Fig.~\ref{phdiaQ1}. 
The additional topological transition is irrelevant for the breaking of the 
global symmetries: the global SU($N_f$)
symmetry is broken in phases $2a$, $2b$, $4a$,
and $4b$, while the U(1)/${\mathbb Z}_q$ symmetry is broken in both phases 
$4a$ and $4b$. 

For $q = 4$ and $Q = 2$ the enlarged symmetry forbids the presence of phases 
$2a$ and $2b$. The only difference with the case $Q=1$ is the presence 
of the topological transition line that divides phase 4 in two different phases.
The behavior of the phase diagram as $q$ increases can be easily inferred.
Since phases $4a$, $4b$ and 3 are not present in the U(1) gauge model, 
the multicritical point 
$M_1$ should move towards larger $\beta$ values, while the multicritical 
point $M_3$ should move towards $\kappa = \infty$. 
For $q=\infty$ continuous 
transitions controlled by the charged fixed point are observed on the 
line starting at $M_2$ and ending at $\kappa=\infty$. Thus, if this transition
survives for finite values of $q$, it should be observed on the line connecting
$M_2$ with $M_3$, which separates
phase 1 from phase $2b$.

\subsection{Critical behavior for $q=10$}

\begin{figure}
\begin{center}
\includegraphics*[width=6cm,angle=-90]{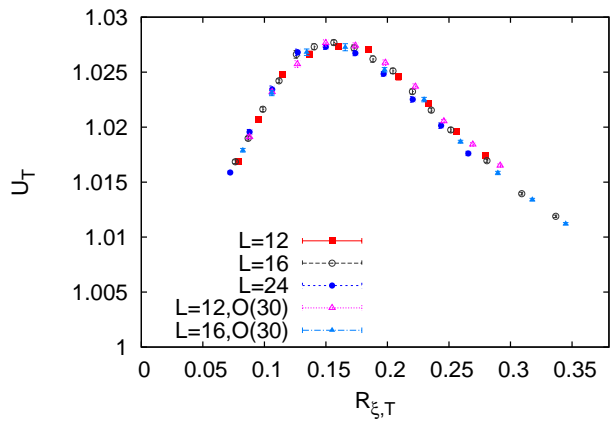}
\end{center}
\caption{Plot of $U_T$ versus $R_{\xi,5}$ for the model with 
$N_f = 15$, $Q =2$, and $q = 10$. Results have been obtained 
varying $\beta$ along the line
$\kappa = 10$. The results are compared with analogous data 
for the O(30) vector model. 
}
\label{UTRxiT-N15-k10}
\end{figure}

\begin{figure}
\includegraphics*[width=5.5cm,angle=-90]{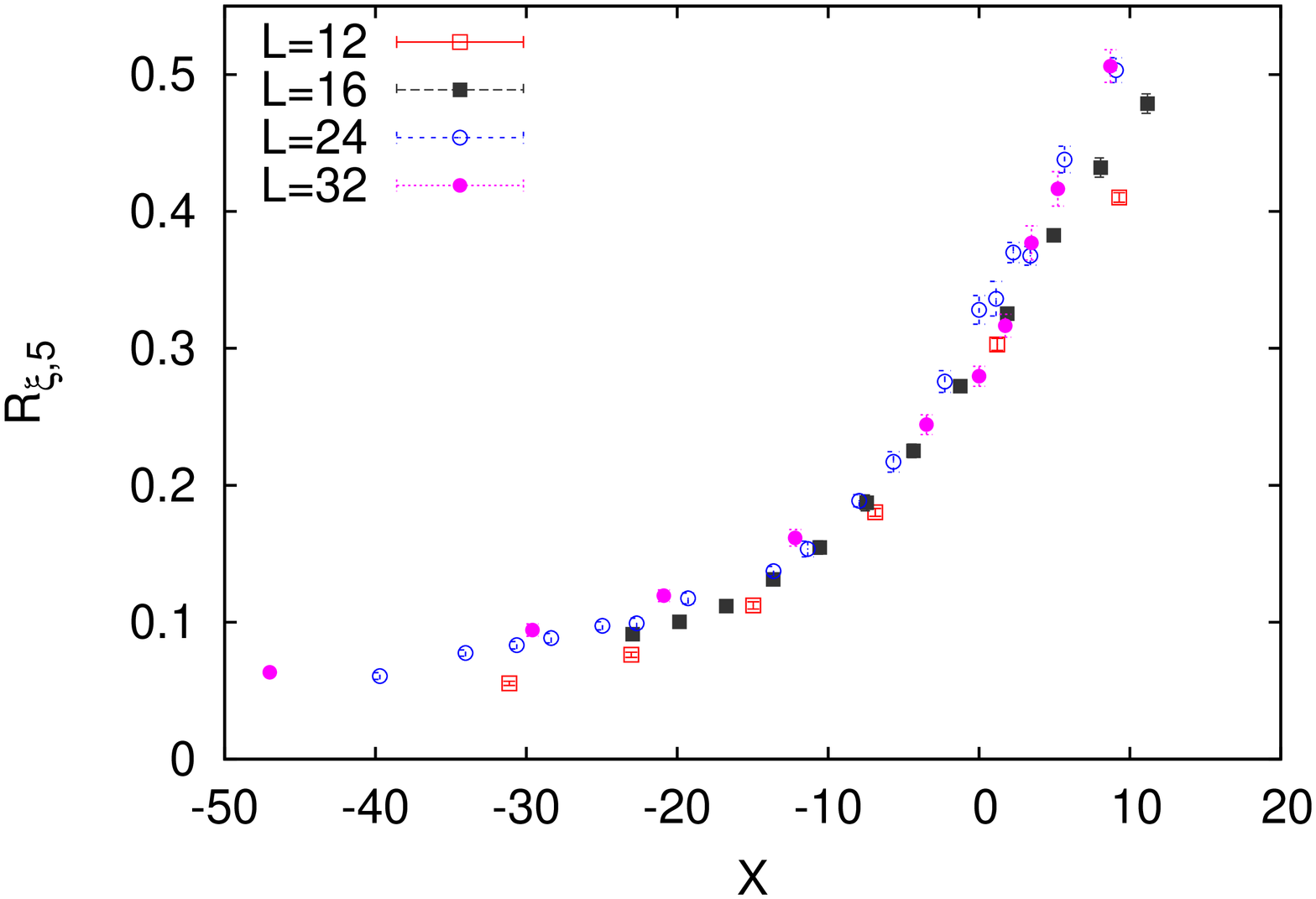}
\includegraphics*[width=5.5cm,angle=-90]{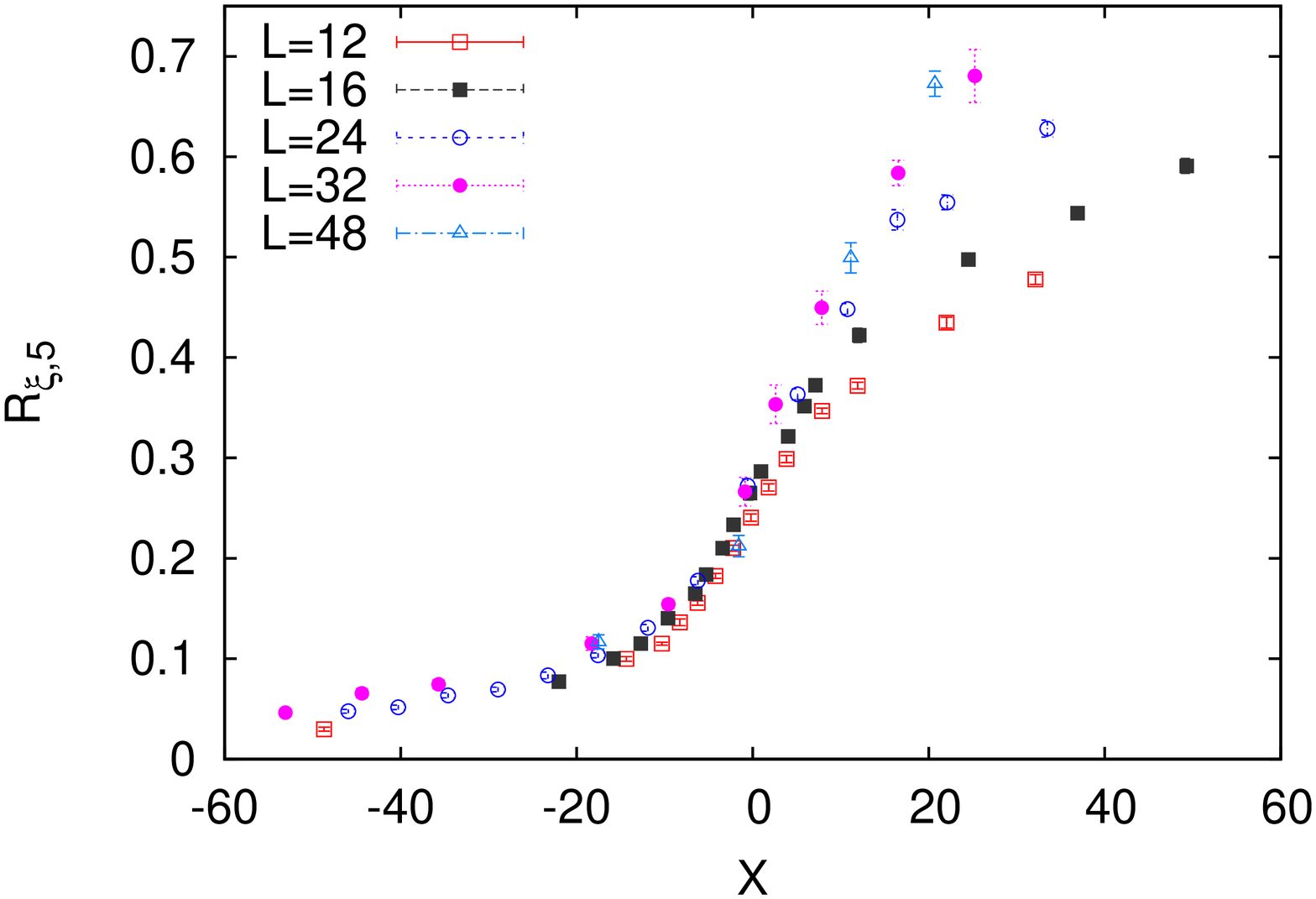}
\caption{Plots of $R_{\xi,5}$ versus $X = (\beta - \beta_c) L^{1/\nu}$
 for the model with $N_f = 15$, $Q =2$, and $q = 10$. 
Results have been obtained varying $\beta$ along the line
$\kappa = 0.4$ (left) and $\kappa = 1.2$ (right). 
We set $\nu = 0.6717$ (the XY-model value \cite{Hasenbusch-19}),
$\beta_c = 14.57$ ($\kappa = 0.4$) and $\beta_c = 13.21$ ($\kappa = 1.2$).
}
\label{Rxi5vsX-N15-k0p4-k1p2}
\end{figure}

For $q=10$, we have first performed runs along lines 
with fixed $\kappa$. We have chosen three different values $\kappa_i$, 
and we have 
performed runs varying $\beta$ at fixed
$\kappa=\kappa_i$. The values $\kappa_i$ have been chosen such as to probe the 
transitions $1\to 2a\to 4a$ (runs at fixed $\kappa_1$), 
the transitions $1\to 2b\to 4b$ (runs at fixed $\kappa_2$), 
and the transitions $3\to 4b$ (runs at fixed $\kappa_3$),
see the right panel of Fig.~\ref{phdiaQ2}. The appropriate $\kappa_i$ values 
have been determined by looking at the behavior for $\beta=0$ and $\infty$
(in these two cases the system corresponds to a pure gauge model,
as discussed in Sec.~\ref{sec4}).
For $\beta = \infty$, the ${\mathbb Z}_2$ topological 
transition line starts at $\kappa \approx 0.76$, while, for $\beta = 0$,
the ${\mathbb Z}_{10}$ topological 
transition line starts \cite{BCCGPS-14} at $\kappa \approx 7.86$. 
Therefore, we 
have chosen $\kappa_1 = 0.4$, $\kappa_2 = 1.2$, and $\kappa_3 = 10$.
For the runs at $\kappa=\kappa_1$ and $\kappa_2$ we have used periodic 
boundary conditions, while open boundary conditions have been used at
$\kappa_3$.  

For $\kappa = 10$, a single transition is observed as $\beta$ is increased,
as expected. To confirm that the transition belongs to the O(30) universality 
class we have determined the universal scaling curve of $U_T$ versus
$R_{\xi,T}$. The data are presented in Fig.~\ref{UTRxiT-N15-k10}, together
with the analogous data for the O(30) vector model. The results for the 
gauge model and for the vector model fall on top of each other, confirming 
that the transition is in the O(30) universality class. 

For both $\kappa = 0.4$ and $\kappa = 1.2$, we 
find two different transitions as $\beta$ is increased. 
For $\kappa = 0.4$ we observe a very strong first-order transition for 
$\beta \approx 11$. The energy shows a strong hysteresis for 
$10.9\lesssim \beta \lesssim 11$ and $11 \lesssim \beta \lesssim 11.3$
for $L = 6$ and 8 (for each $\beta$ we perform 
runs of a few million iterations). 
Equilibrated results are only obtained on lattices of 
size $L=4$. At the transition the SU($N_f$) symmetry is broken---
$\xi_T$ is large and increases rapidly with $L$ on
the large-$\beta$ side of the transition---while the 
U(1)/${\mathbb Z}_q$ symmetry is preserved---the correlation length
$\xi_5$ is small and independent of $L$ at the transition. At 
the second transition,  observed at $\beta \approx 14.6$, $\xi_5$ diverges,
signalling the breaking of the $U(1)/{\mathbb Z}_q$ global symmetry. 
We expect the transition to belong to the XY universality class. 
In Fig.~\ref{Rxi5vsX-N15-k0p4-k1p2} (left panel) we plot $R_{\xi,5}$ versus
$X = (\beta - \beta_c) L^{1/\nu}$ using the XY estimate \cite{Hasenbusch-19}
for the critical exponent $\nu$, $\nu = 0.6717$. 
We observe a reasonable agreement, indicating that 
data are indeed consistent the XY behavior.

\begin{figure}
\begin{center}
\includegraphics*[width=6cm,angle=-90]{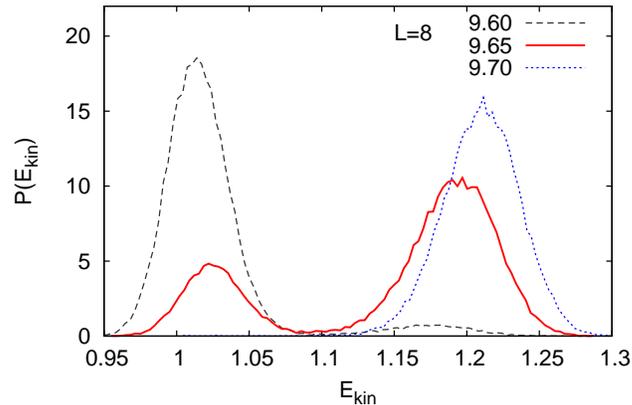}
\end{center}
\caption{Plot of the probability distribution of $E_{\rm kin}$ 
for the model with 
$N_f = 15$, $Q =2$, and $q = 10$. Results have been obtained 
for $\kappa = 1.2$ and a few values of $\beta$ (they are reported in
the legend in the figure).
}
\label{Bimodal-N15-k1p2}
\end{figure}

\begin{figure}
\begin{center}
\includegraphics*[width=6cm,angle=-90]{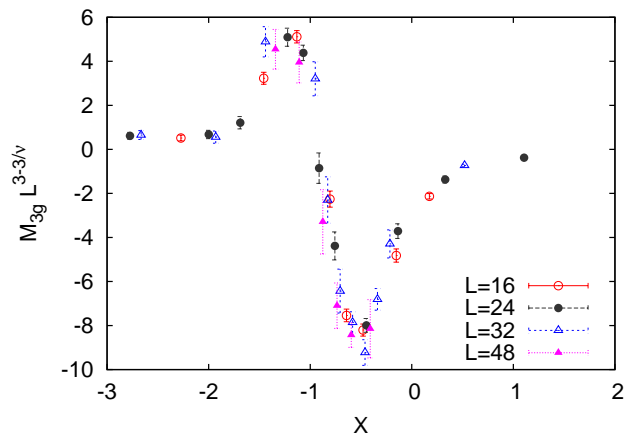}
\end{center}
\caption{Scaling plot of $M_{3g} L^{3 - 3/\nu}$ versus 
$X = (\kappa-\kappa_c) L^{1/\nu}$ for the model with
$N_f = 15$, $Q =2$, and $q = 10$. Results have been obtained
varying $\kappa$ at fixed $\beta = 13$. We set $\nu = 0.629971$
(the Ising model value, see Ref.~\cite{KPSV-16}) and $\kappa_c = 0.817878$.
}
\label{M3g-N15-b13}
\end{figure}

The same analysis has been repeated for $\kappa = 1.2$. As $\beta$
increases, one first observes a transition at $\beta \approx 10$. The
transition is of first order. The scalar-field energy $E_{\rm
kin}$ shows a clear bimodal structure already for $L = 8$, see 
Fig.~\ref{Bimodal-N15-k1p2}. The transition has the same features observed at
$\kappa = 0.4$, indicating that it is associated with the breaking of the 
SU($N_f$) global symmetry. 
As $\beta$ is further increased, a second transition occurs at $\beta
\approx 13.2$, where the correlations $\xi_5$ diverges. 
The numerical data are again consistent with an XY behavior, see 
the right panel of Fig.~\ref{Rxi5vsX-N15-k0p4-k1p2}.

Finally, to obtain an unambiguous check that the runs 
at $\kappa = 1.2$ are indeed probing the transitions $1\to 2b\to 4b$, we have 
performed runs at fixed $\beta = 13$, varying $\kappa$. This value 
of $\beta$ has been chosen on the basis of the results for $\kappa = 1.2$.
Since the XY transition is located at $\beta\approx 13.2$, if the runs 
at $\kappa = 1.2$ were probing the transitions $1\to 2a \to 4a$, 
we would find the topological ${\mathbb Z}_2$ transition at $\kappa  >
1.2$ when varying $\beta$ (see Fig.~\ref{phdiaQ2}). Instead, 
starting from $\kappa =
1.2$, we find a transition at a smaller value of 
$\kappa$, $\kappa \approx 0.81$, confirming the correct identification of 
the transition lines for $\kappa = 1.2$.
The transition at $\kappa \approx 0.81$ has a clear topological nature: in the 
two coexisting phases, 
the SU($N_f$) symmetry is broken 
($U_T\approx 1$ across the transition), while the U(1) scalar modes are
always disordered, $\xi_5\lesssim 1$ in the transition region. 
Clearly, we are observing the 
transition line that separates phases $2a$ and $2b$. Since the transition
is associated with the gauge modes, there are no local order parameters. 
Therefore, to characterize the critical behavior, 
we consider moments of the gauge energy. In particular, we consider the 
third cumulant \cite{SSNHS-03}
\begin{equation} 
M_{3g} = - {1\over V g^3} 
    \left\langle (H_{\rm g} - \langle H_{\rm g}\rangle)^3
    \right\rangle .
\end{equation}
If the transition is continuous, this quantity should scale as 
\begin{equation}
    M_{3g} = L^{3/\nu - 3} f(X) \qquad X = (\kappa - \kappa_c) L^{1/\nu}.
\end{equation}
If we take $\kappa_c = 0.8179$ and we use the Ising estimate \cite{KPSV-16}
$\nu = 0.629971$, data scale nicely, see Fig.~\ref{M3g-N15-b13}.
The transition is therefore continuous in the Ising
universality class. 

The results of the runs at $\beta = 13$ confirm that the runs at $\kappa =
1.2 > \kappa_c(\beta=13) = 0.8$ are indeed probing the transition lines 
$1\to 2b \to 4b$. Since transitions along the 1-$2b$ line 
are of first order,
there is no evidence of the charged universality class. At variance with what 
happens with the global symmetry group---systems with ${\mathbb Z}_q$ symmetry
behave as U(1) systems if $q\ge 4$---in the case of 
the U(1) gauge symmetry, no gauge 
symmetry enlargement occurs: the charged universality class can apparently
be observed only for the U(1) model, i.e., for $q=\infty$.

The runs at $\beta = 13$ also show that, along the line running from 
$\beta = \infty$ up to the multicritical point $M_2$, 
the transitions are always 
topological in the Ising universality class. Scalar fields 
play no role, so that the same critical behavior is observed on the whole
line starting at $\beta=\infty$ and ending at the multicritical point 
$M_2$, see Fig.~\ref{phdiaQ2}. 

\subsection{Phase behavior for $q=4$ and $q=6$}

\begin{figure}
\includegraphics*[width=5.5cm,angle=-90]{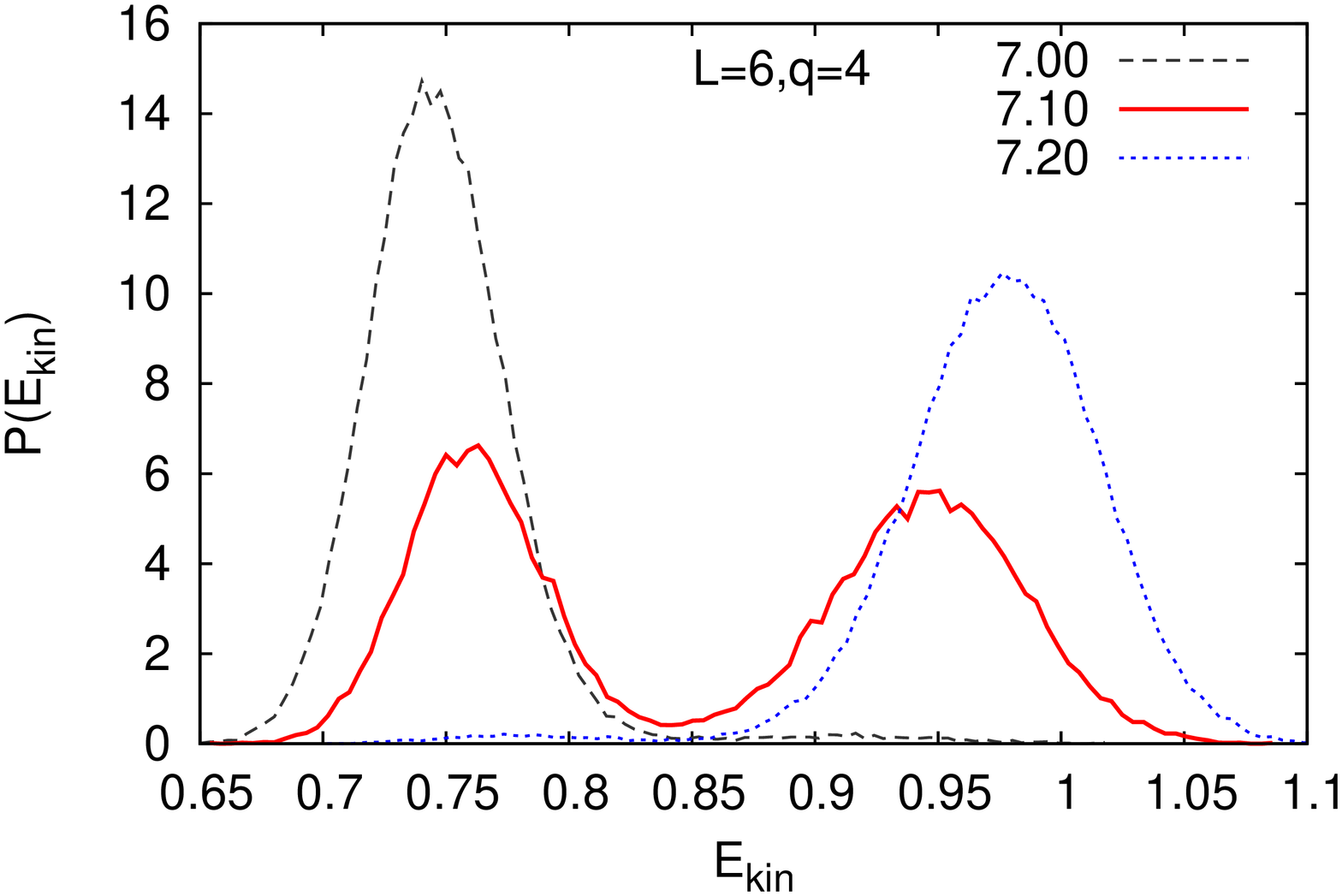}
\includegraphics*[width=5.5cm,angle=-90]{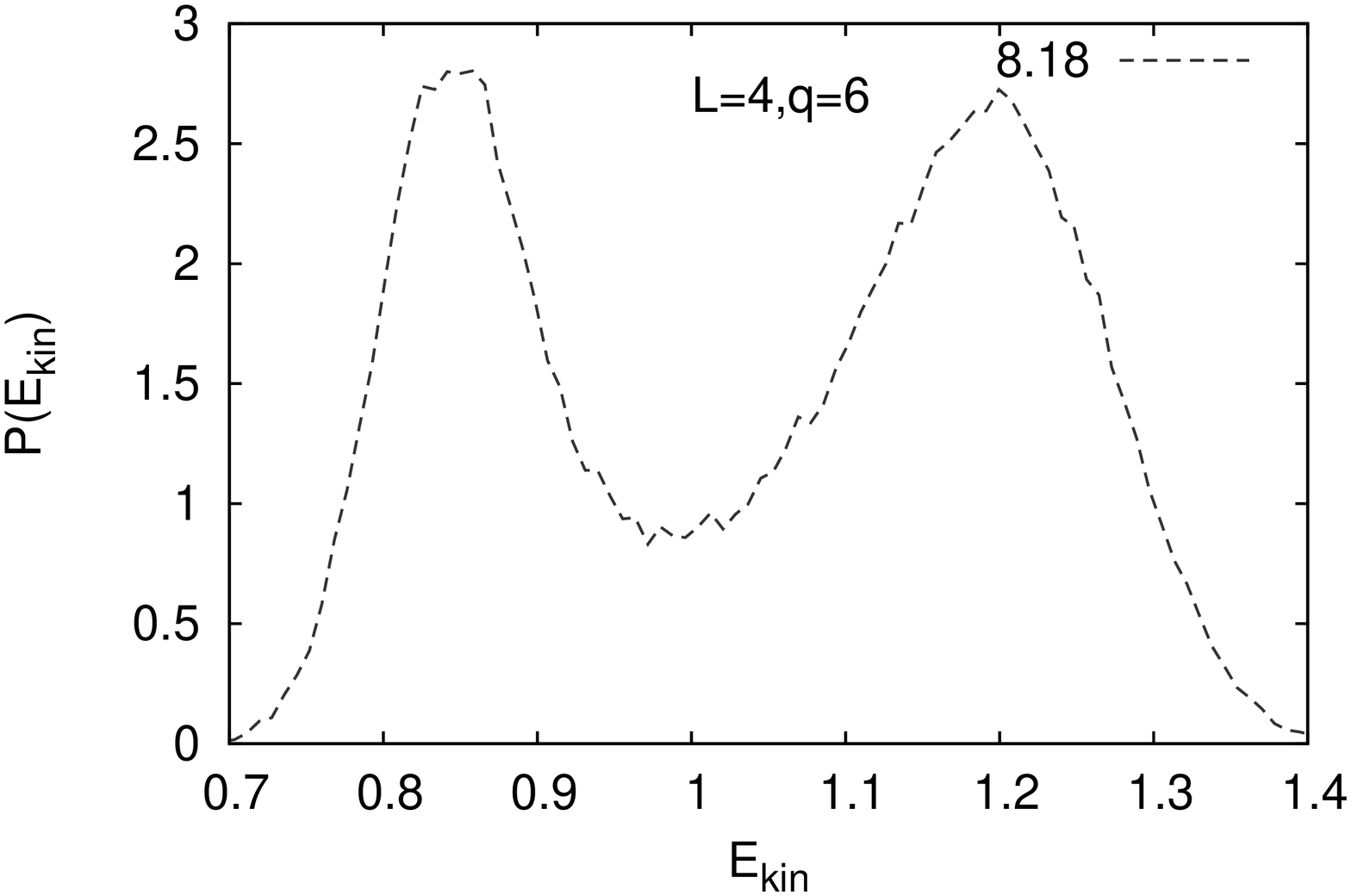}
\caption{Plot of the probability distribution of $E_{\rm kin}$ 
for the model with $N_f = 15$ and $Q =2$, for $\kappa = 1.2$ 
and different values of $\beta$ (they are reported in the legend). 
Results for $q = 4$ and $q=6$ are reported in the left and 
right panel, respectively. Periodic boundary conditions are used.
}
\label{Bimodal-N15-q4-q6}
\end{figure}

Beside the extensive simulations for $q=10$, we also performed some
simulations for $q=4$ and $q=6$. We chose $\kappa = 1.2$ in both cases 
and varied $\beta$, in order to identify the transitions 1-$2b$, $2b$-$4b$
for $q=6$ and the single transition 1-$4b$ for $q=4$ (the enlarged O(30) global
symmetry forbids phases $2a$ and $2b$),
see Fig.~\ref{phdiaQ2}. 
For $q=4$ numerical simulations confirm the existence of 
a single transition. Using 
both open and periodic boundary conditions, we verify 
that the transition is of
first order: the energy has a bimodal distribution already on small lattices, 
see the left panel of Fig.~\ref{Bimodal-N15-q4-q6}. 

For $q=6$ we would expect the same phase diagram as that shown in
Fig.~\ref{phdiaQ2}. Surprisingly, our data at fixed $\kappa =1.2$ 
are consistent with a single first-order transition.
The latent heat is large---the distribution of the energy is bimodal 
already for $L=4$, see the right panel of Fig.~\ref{Bimodal-N15-q4-q6}---and 
we are not able to obtain equilibrated results for $L\ge 6$. On the
large-$\beta$ side of the transition, both $\xi_T$ and $\xi_3$ increase 
rapidly with $L$ (comparing data for $L=8$ and 16, we estimate an effective
behavior $\xi\sim L^{1.6}$), which would apparently indicate that both 
the SU($N_f$) and the U(1)/${\mathbb Z}_6$ symmetry are broken. 
We do not see any reason why the model with $q=6$ should have 
a phase diagram different from that of the model with $q=10$, 
presented in Fig.~\ref{phdiaQ2}. A possible explanation of the 
observed behavior that is consistent with the general picture 
obtained from the simulations with $q=10$ is the following.
It is possible that 
$\xi_3$ on the large-$\beta$ side of the transition 
is finite (as expected in
phase $2b$), but large ($\xi_3 \gg 10$). 
In this case, our numerical simulations on lattices of 
size of order 10 would be unable to distinguish phase $2b$ from phase $4b$.

\section{Conclusions} \label{sec7}

In this paper we consider a variant of the usual 
three-dimensional compact AH model in which 
a complex $N_f$ component scalar field of charge $Q$ is coupled with 
a U(1) Abelian gauge field. Here, gauge fields are phases that take 
only a set of $q$ discrete values so that the model is 
${\mathbb Z}_q$ gauge invariant.
Because of the smaller gauge group, the theory is globally invariant under 
a larger set of symmetry transformations. Beside SU($N_f)$ transformations,
there is also a nontrivial 
invariance under global U(1)/${\mathbb Z}_q$ phase changes 
of the scalar fields. The larger invariance group allows for a 
richer phase diagram, with additional transition lines associated with
the breaking of the U(1)/${\mathbb Z}_q$ symmetry. 

By means of numerical Monte Carlo simulations,
we determine the phase diagram and investigate the nature of the different
transition lines in two different cases. First, we consider the model with 
charge-1 two-component scalar fields; second, we study the 
model with scalar fields of charge $Q=2$ and 
$N_f = 15$ components. These two sets of parameters have been chosen
on the basis of the results for the U(1) gauge AH model \cite{BPV-22-AHq},
which show that 
in these two cases gauge fields play a completely different role.

For U(1) gauge fields,
the transitions observed for $N_f=2$ and $Q=1$ are uniquely
characterized by the critical fluctuations of the scalar fields. 
Gauge fields are only relevant in selecting the physical observables (i.e.,
gauge-invariant quantities), but do not determine the nature of 
the critical behavior, which can be predicted by using an 
effective LGW theory 
for a gauge-invariant order parameter, with no explicit gauge fields. 
Therefore, the universality class of the transition depends only
on the transformation properties of the order parameter under 
the global SU(2) symmetry and 
the symmetry breaking pattern at the transition. 
In the SO(3) language [note that SU(2)$/{\mathbb Z}_2$ = SO(3)], the order 
parameter is an SO(3) vector and the symmetry breaking pattern corresponds to 
$SO(3)\to SO(2)$. Therefore, one predicts the transition to belong to the SO(3)
vector universality class, a result that is fully supported by
simulations (see, e.g.,~\cite{PV-19-AH3d}).
In the discrete gauge model, along the line 
where the SU(2) symmetry is broken, 
we observe the same SO(3) vector behavior for any $q\ge 3$. 
This result further confirms correctness of the effective 
LGW effective theory and the irrelevance of the gauge fields along
the O(3) transition line.

As we already mentioned, the discrete model is also invariant
under the U(1)$/{\mathbb Z}_q$ group. 
We find transition lines where this Abelian global symmetry is
broken. They occur within the SU(2) ordered 
phase. Again, gauge fields are irrelevant for the determination of the 
critical behavior: transitions belong to the XY universality class, 
as in systems without gauge fields. We also observe the presence of lines where 
both the SU(2) group and the U(1)$/{\mathbb Z}_q$ group break
simulteneuosly. On these lines 
the critical behavior belongs to the O(4) 
universality class. We observe here an enlargement of the global symmetry
group: the O(4) group is the smallest simple group that has U(2) has one of its
subgroups. The symmetry enlargement is strictly related to the 
irrelevance of the ${\mathbb Z}_q$ gauge interactions, as the model we consider 
is O(4) invariant in the absence of gauge fields: the addition
of ${\mathbb Z}_q$ gauge fields represents an irrelevant perturbation
(in the renormalization-group sense) of the O(4) vector fixed point.

We have then considered the model with $Q=2$ and $N_f=15$.
In the compact U(1) gauge model, for these values of the parameters,
there is a transition line where the SU($N_f$) symmetry is broken
and transitions belong to the {\em charged} AH universality class
\cite{BPV-20-AHq2,BPV-22-AHq}. Gauge fields 
are crucial at the transition and the effective field theory model that 
describes the universal features of the transition is the continuum
AH model with explicit gauge fields. 
We have studied whether these charged transitions also occur when
the U(1) group is replaced by one of its ${\mathbb Z}_q$ subgroups, i.e., 
if there is a {\rm gauge symmetry enlargement}, as it occurs for global
U(1) symmetries. Our numerical Monte Carlo results for $q=6$ and 10
give a negative answer. 
In the compact models with discrete gauge groups that we consider,
the transitions along the line where 
the SU($N_f$) symmetry is broken are always of first order. Apparently, 
in the compact AH model it is
crucial that the microscopic model is U(1) gauge invariant to observe 
the charged behavior, at variance with what happens along the {\em uncharged}
transition lines, where the exact gauge symmetry plays little role.

\bigskip

Simulations were performed at the INFN Computation Center  of the 
INFN, Sezione di Roma.

\end{document}